\documentclass[acmsmall,screen]{acmart}

\AtBeginDocument{%
  }

\setcopyright{none}
\copyrightyear{2018}
\acmYear{2018}
\acmDOI{XXXXXXX.XXXXXXX}

\acmConference[\red{Preprint}]{Make sure to enter the correct
  conference title from your rights confirmation emai}{January 28,
  2024}{Canada}
\acmPrice{15.00}
\acmISBN{978-1-4503-XXXX-X/18/06}

\usepackage{algorithmic}
\usepackage{amsmath}
\usepackage{amsfonts}
\usepackage{booktabs}
\usepackage{colortbl}
\usepackage[inline]{enumitem}
\usepackage{graphicx}
\usepackage{hyperref}
\usepackage{listings}
\usepackage{makecell}
\usepackage{mathtools}
\usepackage{multirow}
\usepackage{nameref}
\usepackage{namedef}
\usepackage{pbox}
\usepackage{pgffor}
\usepackage{soul}
\usepackage{tabularx}
\usepackage{textcomp}
\usepackage{varioref}
\usepackage{xcolor}
\usepackage{xspace}
\usepackage{xstring}

\definecolor{backcolour}{rgb}{0.95,0.95,0.92}

\newcolumntype{Y}{>{\centering\arraybackslash}X}
\newcolumntype{Z}{>{\rightline\arraybackslash}X}

\def\SALMLabeledMigCount{310\xspace}
\def\SALMLabeledRepoCount{298\xspace}
\def\SALMLabeledLPCount{131\xspace}
\def\SALMLabeledLibCount{193\xspace}
\def\SALMLabeledDomainCount{25\xspace}
\def\TotalLabeledMigCount{385\xspace}
\def\TotalLabeledRepoCount{355\xspace}
\def\TotalLabeledLPCount{159\xspace}
\def\TotalLabeledLibCount{229\xspace}
\def\TotalLabeledDomainCount{36\xspace}
\def\CandidateSalmMigs{1,269\xspace}
\def\LibPairsWithSameImportNameCount{2\xspace}

\def\VOneMigCount{75\xspace}
\def\VOneRepoCount{57\xspace}
\def\VOneLPCount{34\xspace}
\def\VOneLibCount{55\xspace}
\def\VOneDomainCount{11\xspace}
\def\VTwoCCCount{3,096\xspace}
\def\VTwoMigCount{335\xspace}
\def\VTwoRepoCount{311\xspace}
\def\VTwoLPCount{141\xspace}
\def\VTwoLibCount{208\xspace}
\def\VTwoDomainCount{35\xspace}
\def\RoundOneMigCount{13\xspace}
\def\RoundTwoMigCount{19\xspace}
\def\RoundThreeMigCount{20\xspace}
\def\RoundFourMigCount{259\xspace}

\def\RoundFourCCCount{1,233\xspace}
\def\RoundOneRepoCount{13\xspace}
\def\RoundTwoRepoCount{19\xspace}
\def\RoundThreeRepoCount{20\xspace}
\def\RoundFourRepoCount{249\xspace}
\def\RoundOneLPCount{12\xspace}
\def\RoundTwoLPCount{14\xspace}
\def\RoundThreeLPCount{12\xspace}
\def\RoundFourLPCount{116\xspace}
\def\RoundOneLibCount{21\xspace}
\def\RoundTwoLibCount{23\xspace}
\def\RoundThreeLibCount{24\xspace}
\def\RoundFourLibCount{174\xspace}
\def\RoundOneDomainCount{3\xspace}
\def\RoundTwoDomainCount{7\xspace}
\def\RoundThreeDomainCount{9\xspace}
\def\RoundFourDomainCount{24\xspace}
\def\RequiredSALMSampleSize{296\xspace}
\def\ActualSALMSampleSize{311\xspace}
\def\SALMLabeledDomainPercent{100\%\xspace}
\def\SALMLabeledLPPercent{66\%\xspace}
\def\SALMTotalMigCount{1,559\xspace}
\def\SALMTotalRepoCount{1,291\xspace}
\def\SALMTotalLPCount{199\xspace}
\def\SALMTotalLibCount{265\xspace}
\def\SALMTotalDomainCount{25\xspace}
\def\GrandTotalMigCount{1,634\xspace}
\def\GrandTotalRepoCount{1,338\xspace}
\def\GrandTotalLPCount{224\xspace}
\def\GrandTotalLibCount{296\xspace}
\def\GrandTotalDomainCount{36\xspace}
\def\FunctionCallFunctionCallLPCount{103\xspace}
\def\FunctionCallFunctionCallLPPercent{73\%\xspace}

\def\FunctionCallFunctionCallArgumentAdditionPercent{28\%\xspace}

\def\FunctionCallFunctionCallArgumentTransformationPercent{29\%\xspace}

\def\FunctionCallAttributeLPCount{16\xspace}
\def\FunctionCallAttributeLPPercent{11\%\xspace}

\def\FunctionCallDecoratorLPCount{3\xspace}
\def\FunctionCallDecoratorLPPercent{2\%\xspace}

\def\FunctionCallDecoratorArgumentAdditionPercent{50\%\xspace}

\def\FunctionCallNoneLPCount{24\xspace}
\def\FunctionCallNoneLPPercent{17\%\xspace}

\def\FunctionCallTotalLPCount{109\xspace}
\def\FunctionCallTotalLPPercent{77\%\xspace}

\def\FunctionCallTotalArgumentTransformationPercent{28\%\xspace}

\def\AttributeFunctionCallLPCount{9\xspace}
\def\AttributeFunctionCallLPPercent{6\%\xspace}

\def\AttributeAttributeLPCount{18\xspace}
\def\AttributeAttributeLPPercent{13\%\xspace}

\def\AttributeNoneLPCount{3\xspace}
\def\AttributeNoneLPPercent{2\%\xspace}

\def\AttributeTotalLPCount{25\xspace}
\def\AttributeTotalLPPercent{18\%\xspace}

\def\DecoratorFunctionCallLPCount{1\xspace}
\def\DecoratorFunctionCallLPPercent{0.7\%\xspace}

\def\DecoratorFunctionCallArgumentAdditionPercent{100\%\xspace}

\def\DecoratorDecoratorLPCount{8\xspace}
\def\DecoratorDecoratorLPPercent{6\%\xspace}

\def\DecoratorDecoratorArgumentAdditionPercent{20\%\xspace}

\def\DecoratorDecoratorArgumentNameChangePercent{30\%\xspace}

\def\DecoratorNoneLPCount{2\xspace}
\def\DecoratorNoneLPPercent{1\%\xspace}

\def\DecoratorTotalLPCount{9\xspace}
\def\DecoratorTotalLPPercent{6\%\xspace}

\def\FunctionReferenceFunctionReferenceLPCount{3\xspace}
\def\FunctionReferenceFunctionReferenceLPPercent{2\%\xspace}

\def\FunctionReferenceNoneLPCount{1\xspace}
\def\FunctionReferenceNoneLPPercent{0.7\%\xspace}

\def\FunctionReferenceTotalLPCount{4\xspace}
\def\FunctionReferenceTotalLPPercent{3\%\xspace}

\def\TypeTypeLPCount{10\xspace}
\def\TypeTypeLPPercent{7\%\xspace}

\def\TypeNoneLPCount{1\xspace}
\def\TypeNoneLPPercent{0.7\%\xspace}

\def\TypeTotalLPCount{10\xspace}
\def\TypeTotalLPPercent{7\%\xspace}

\def\ExceptionExceptionLPCount{11\xspace}
\def\ExceptionExceptionLPPercent{8\%\xspace}

\def\ExceptionNoneLPCount{4\xspace}
\def\ExceptionNoneLPPercent{3\%\xspace}

\def\ExceptionTotalLPCount{14\xspace}
\def\ExceptionTotalLPPercent{10\%\xspace}

\def\ImportImportLPCount{137\xspace}
\def\ImportImportLPPercent{97\%\xspace}

\def\ImportNoneLPCount{9\xspace}
\def\ImportNoneLPPercent{6\%\xspace}

\def\ImportTotalLPCount{138\xspace}
\def\ImportTotalLPPercent{98\%\xspace}

\def\NoneFunctionCallLPCount{15\xspace}
\def\NoneFunctionCallLPPercent{11\%\xspace}

\def\NoneAttributeLPCount{5\xspace}
\def\NoneAttributeLPPercent{4\%\xspace}

\def\NoneDecoratorLPCount{4\xspace}
\def\NoneDecoratorLPPercent{3\%\xspace}

\def\NoneFunctionReferenceLPCount{1\xspace}
\def\NoneFunctionReferenceLPPercent{0.7\%\xspace}

\def\NoneTypeLPCount{3\xspace}
\def\NoneTypeLPPercent{2\%\xspace}

\def\NoneExceptionLPCount{3\xspace}
\def\NoneExceptionLPPercent{2\%\xspace}

\def\NoneImportLPCount{8\xspace}
\def\NoneImportLPPercent{6\%\xspace}

\def\NoneTotalLPCount{23\xspace}
\def\NoneTotalLPPercent{16\%\xspace}

\def\TotalFunctionCallLPCount{107\xspace}
\def\TotalFunctionCallLPPercent{76\%\xspace}

\def\TotalAttributeLPCount{30\xspace}
\def\TotalAttributeLPPercent{21\%\xspace}

\def\TotalDecoratorLPCount{12\xspace}
\def\TotalDecoratorLPPercent{9\%\xspace}

\def\TotalDecoratorArgumentDeletionPercent{67\%\xspace}

\def\TotalFunctionReferenceLPCount{4\xspace}
\def\TotalFunctionReferenceLPPercent{3\%\xspace}

\def\TotalTypeLPCount{11\xspace}
\def\TotalTypeLPPercent{8\%\xspace}

\def\TotalExceptionLPCount{12\xspace}
\def\TotalExceptionLPPercent{9\%\xspace}

\def\TotalImportLPCount{137\xspace}
\def\TotalImportLPPercent{97\%\xspace}

\def\TotalNoneLPCount{31\xspace}
\def\TotalNoneLPPercent{22\%\xspace}

\def\TotalLPCount{141\xspace}
\def\TotalLPPercent{100\%\xspace}
\def\TotalElementNameChangePercent{65\%\xspace}
\def\TotalArgumentAdditionPercent{22\%\xspace}
\def\TotalArgumentDeletionPercent{18\%\xspace}

\def\TotalAsyncTransformationPercent{4\%\xspace}
\def\TotalOutputTransformationPercent{7\%\xspace}
\def\TotalParameterAdditionToDecoratedFunctionPercent{2\%\xspace}

\def\FunctionCallFunctionCallSameNamePercent{32\%\xspace}
\def\HasNonFunctionLPPercent{40\%\xspace}

\def\MigLOCMedian{8\xspace}

\def\MigLOCMax{758\xspace}

\def\NumAPIsMedian{7\xspace}

\def\NumChangesMedian{3\xspace}

\def\UniqueAPIsMedian{4\xspace}

\def\UniqueAPIsMax{43\xspace}

\def\UniqueMappingsMedian{2\xspace}

\def\UniqueMappingsMax{22\xspace}

\def\UniqueCCCombo{167\xspace}
\def\HasFcComboCount{152\xspace}
\def\HasFcComboPercent{91\%\xspace}
\def\ComboFunctionCallWithElemNCPercent{8.2\%\xspace}

\def\ComboJustFcMigPercent{56\%\xspace}
\def\ComboJustNonFCMigPercent{7\%\xspace}
\def\ComboMixFcNonFcMigPercent{37\%\xspace}
\def\ComboHasNonFCMigPercent{44\%\xspace}
\def\AttributeWithFCInDiffCCPercent{13\%\xspace}
\def\AttributeWithFCInSameCCPercent{9\%\xspace}
\def\DecoratorWithFCInDiffCCPercent{11\%\xspace}
\def\DecoratorWithFCInSameCCPercent{8\%\xspace}
\def\ExceptionWithFCInSameCCPercent{5\%\xspace}
\def\TypeWithFCInSameCCPercent{3.5\%\xspace}
\def\FuncRefWithFCInSameCCPercent{1.6\%\xspace}

\def\diffWidth{1}
\newcommand{\red}[1]{\textcolor{red}{#1}}

\newcommand{\sn}[1]{\textcolor{purple}{\textbf{Sarah:}~#1}\xspace}

\newcommand{\hide}[1]{\ignorespaces}

\newcommand{\migbench}{\textsc{PyMigBench}\xspace}
\newcommand{\migbenchTwo}{\migbench-2.0\xspace}
\newcommand{\taxonomy}{\textsc{PyMigTax}\xspace}
\newcommand{\salm}{SALM\xspace}
\newcommand{\etal}{et al.\xspace}

\newcommand{\ie}{i.e.\xspace}

\newcommand{\artifactURL}{\url{https://doi.org/10.6084/m9.figshare.24216858.v2}}

\newcommand{\libmigs}[0]{library migrations\xspace}

\newcommand{\ccs}[0]{code changes\xspace}
\newcommand{\cc}[0]{code change\xspace}
\newcommand{\codechange}[0]{migration-related code change\xspace}

\newcommand{\codechanges}[0]{migration-related code changes\xspace}

\newcommand{\libpairs}[0]{library pairs\xspace}

\newcommand{\fc}[0]{{function call}\xspace}
\newcommand{\fcs}[0]{{function calls}\xspace}

\newcommand{\ca}[0]{{function reference}\xspace}
\newcommand{\Ca}[0]{Function reference\xspace}

\newcommand{\ty}[0]{{type}\xspace}

\newcommand{\im}[0]{{import}\xspace}
\newcommand{\imp}[0]{\im}

\newcommand{\ex}[0]{{exception}\xspace}

\newcommand{\at}[0]{{attribute}\xspace}

\newcommand{\ats}[0]{{attributes}\xspace}
\newcommand{\de}[0]{{decorator}\xspace}

\newcommand{\des}[0]{{decorators}\xspace}

\newcommand{\oo}[0]{one-to-one\xspace}
\newcommand{\om}[0]{one-to-many\xspace}
\newcommand{\mo}[0]{many-to-one\xspace}
\newcommand{\mm}[0]{many-to-many\xspace}
\newcommand{\zo}[0]{zero-to-one\xspace}
\newcommand{\oz}[0]{one-to-zero\xspace}

\newcommand{\argAdd}[0]{{argument addition}\xspace}
\newcommand{\ArgAdd}[0]{{Argument addition}\xspace}
\newcommand{\argDel}[0]{{argument deletion}\xspace}
\newcommand{\argTrans}[0]{{argument transformation}\xspace}
\newcommand{\enc}[0]{{element name change}\xspace}
\newcommand{\asyncChange}[0]{{async transformation}\xspace}
\newcommand{\AsyncChange}[0]{{Async transformation}\xspace}

\newcommand{\outTrans}[0]{{output transformation}\xspace}
\newcommand{\argNC}[0]{{argument name change}\xspace}
\newcommand{\ArgNC}[0]{{Argument name change}\xspace}

\newcommand{\paramAddToDecorate}[0]{{parameter addition to decorated function}\xspace}
\newcommand{\ParamAddToDecorate}[0]{{Parameter addition to decorated function}\xspace}

\newcommand{\argAddShort}[0]{{argAdd}\xspace}
\newcommand{\argDelShort}[0]{{argDel}\xspace}
\newcommand{\argTransShort}[0]{{argTrans}\xspace}
\newcommand{\encShort}[0]{{elemNC}\xspace}
\newcommand{\asyncChangeShort}[0]{{asyncTrans}\xspace}

\newcommand{\outTransShort}[0]{{outTrans}\xspace}

\newcommand{\lib}[1]{\textit{#1}\xspace}

\newcommand{\apimaptechnique}[0]{API mapping\xspace}
\newcommand{\transform}{client code transformation\xspace}

\newcommand{\commitLinkPlain}[2]{\href{https://github.com/#1/commit/#2}{#1\##2}}

\definecolor{rem}{RGB}{255,40,80}
\definecolor{add}{RGB}{20,177,64}
\definecolor{cardinality}{HTML}{0071BC}
\definecolor{spe}{HTML}{F13F99}
\definecolor{tpe}{HTML}{00A64F}
\definecolor{prop}{HTML}{FAA21A}

\newcommand{\bg}[2][yellow]{{%
    \colorlet{foo}{#1}%
    \sethlcolor{foo}\hl{\mbox{#2}}}%
}

\newcommand{\ccDesc}[3]{
    \textit{#1}~$\rightarrow$~\textit{#2} | \textit{#3}
}

\newcommand{\ccDescLong}[4]{
    \textit{#1}~$\xrightarrow{#4}$~\textit{#2} | \textit{#3}
}

\newcommand{\migDesc}[4]{
    \lib{#1} to \lib{#2} migration extracted/adapted from \href{https://github.com/#3/commit/#4}{\textit{#3\##4}}.
}

\newcommand{\remLine}[1]{\bg[rem!10]{$-$#1}}
\newcommand{\addLine}[1]{\bg[add!10]{$+$#1}}
\newcommand{\ccLines}[2]{\bg[rem!10]{#1}:\bg[add!10]{#2}}

\usepackage[absolute,overlay]{textpos}
\usepackage{tikz}
\usepackage{xcolor}
\usepackage{calc}

\newcounter{findingctr}

\newcommand{\finding}[2]{\refstepcounter{findingctr}\emph{#1:\label{box:#2}}}

\newlength{\boxw}
\newlength{\boxh}
\newlength{\shadowsize}
\newlength{\boxroundness}
\newlength{\tmpa}
\newsavebox{\shadowblockbox}

\newenvironment{findingenv}[2]%
{\vspace{0.2cm}\noindent
\begin{lrbox}{
\shadowblockbox
}
\begin{minipage}{.98\columnwidth}
\finding{#1}{#2}~}%
{\end{minipage}\end{lrbox}%
\settowidth{\boxw}{\usebox{\shadowblockbox}}%
\settodepth{\tmpa}{\usebox{\shadowblockbox}}%
\settoheight{\boxh}{\usebox{\shadowblockbox}}%
\addtolength{\boxh}{\tmpa}%
\begin{tikzpicture}
\addtolength{\boxw}{\boxroundness * 2}
\addtolength{\boxh}{\boxroundness * 2}

\foreach \x in {0,.05,...,1}
{
\setlength{\tmpa}{\shadowsize * \real{\x}}
\fill[xshift=\shadowsize - 1pt,yshift=-\shadowsize + 
1pt,black,opacity=.04,rounded corners=\boxroundness] 
(\tmpa, \tmpa) rectangle +(\boxw - \tmpa - \tmpa, \boxh - \tmpa - 
\tmpa);
}

\filldraw[fill=white!50, draw=black!80, rounded corners=\boxroundness] (0, 
0) rectangle (\boxw, \boxh);
\draw node[xshift=\boxroundness,yshift=\boxroundness,inner sep=0pt,outer 
sep=0pt,anchor=south west] (0,0) {\usebox{\shadowblockbox}};
\end{tikzpicture}\vspace{0cm}%
}

\begin{document}

\title{Characterizing Python Library Migrations}

\author{Mohayeminul Islam}
\email{mohayemin@ualberta.ca}
\affiliation{
  \institution{University of Alberta}
  \city{Edmonton}
  \state{Alberta}
  \country{Canada}
}
\author{Ajay Kumar Jha}
\email{ajay.jha.1@ndsu.edu}
\affiliation{%
  \institution{North Dakota State University}
  \city{Fargo}
  \state{North Dakota}
  \country{USA}
}

\author{Ildar Akhmetov}
\email{i.akhmetov@northeastern.edu}
\affiliation{%
  \institution{Northeastern University}
  \city{Vancouver}
  \state{British Columbia}
  \country{Canada}
}

\author{Sarah Nadi}
\email{nadi@ualberta.ca}
\affiliation{%
  \institution{University of Alberta}
  \city{Edmonton}
  \state{Alberta}
  \country{Canada}
}

\renewcommand{\shortauthors}{Islam et al.}

\begin{abstract}
Developers heavily rely on Application Programming Interfaces (APIs) from libraries to build their software.
As software evolves, developers may need to replace the used libraries with alternate libraries, a process known as \textit{library migration}.
Doing this manually can be tedious, time-consuming, and prone to errors.
Automated migration techniques can help alleviate some of this burden.
However, designing effective automated migration techniques requires understanding the types of code changes required to transform client code that used the old library to the new library.
This paper contributes an empirical study that provides a holistic view of Python library migrations, both in terms of the code changes required in a migration and the typical development effort involved.
We manually label \VTwoCCCount \codechanges in \VTwoMigCount Python library migrations from \VTwoRepoCount client repositories spanning \VTwoLPCount library pairs from \VTwoDomainCount domains.
Based on our labeled data, we derive a taxonomy for describing \codechanges, \taxonomy.
Leveraging \taxonomy and our labeled data, we investigate various characteristics of Python library migrations, such as the types of program elements and properties of API mappings, the combinations of types of \codechanges in a migration, and the typical development effort required for a migration.
Our findings highlight various potential shortcomings of current library migration tools.
For example, we find that \HasNonFunctionLPPercent of library pairs have API mappings that involve non-function program elements, while most library migration techniques typically assume that function calls from the source library will map into (one or more) function calls from the target library.
As an approximation for the development effort involved, we find that, on average, a developer needs to learn about \UniqueAPIsMedian APIs and \UniqueMappingsMedian API mappings to perform a migration, and change \MigLOCMedian lines of code.
However, we also found cases of migrations that involve up to \UniqueAPIsMax unique APIs, \UniqueMappingsMax API mappings, and \MigLOCMax lines of code, making them harder to manually implement.
Overall, our contributions provide the necessary knowledge and foundations for developing automated Python library migration techniques. 
We make all data and scripts related to this study publicly available at \artifactURL.

\end{abstract}


\maketitle

\section{Introduction}
\label{sec:intro}
Modern software development heavily relies on third-party libraries~\cite{understandingAPIUsage, empiricalAPIUsage, trivialNPMPackage}, as they can enhance developer efficiency~\cite{softwareReuse}
 and improve the reliability and maintainability of software systems~\cite{reinventingTheWheels}.
However, the libraries that an application depends on may become obsolete over time~\cite{deprecatedPythonAPIs, apiDeprecation, he2021multi}.
Libraries with vulnerabilities or bugs can also adversely impact the applications that use them~\cite{keepMeUpdated, kula2018developers}.
Moreover, developers might discover newer, better-performing, or easier-to-use libraries~\cite{kabinna2016logging, selectingLibraries, he2021multi}.
In all these situations, developers often need to replace a currently used library with an alternative one, a process commonly referred to as \textit{library migration}.

\begin{figure}
    \centering
    \includegraphics[width=\diffWidth\columnwidth]{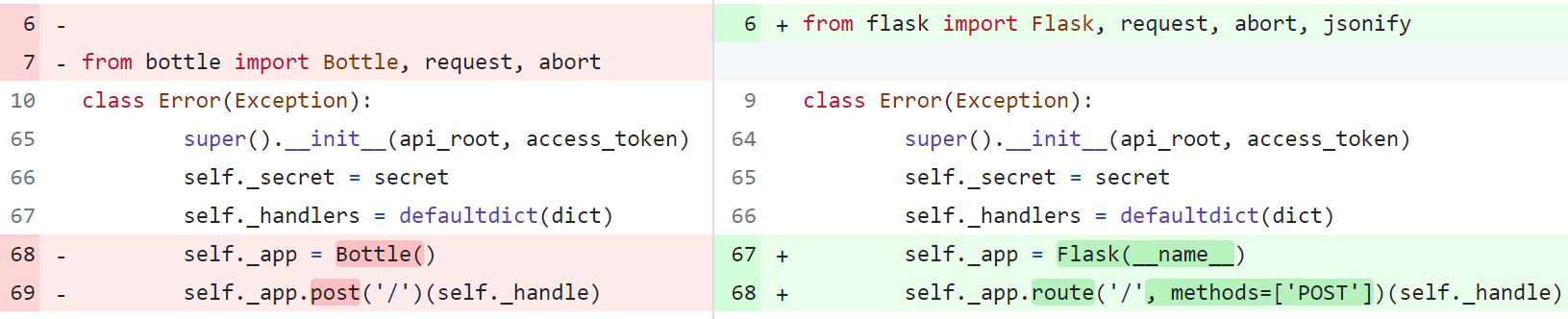}
    \caption{
        \migDesc{bottle}{flask}{cqmoe/python-cqhttp}{f9f083ec}
    }
    \label{fig:f9f083ec}
    \Description{}
\end{figure}

The library migration process typically involves finding which APIs from the new library can replace the used APIs from the old library (\textit{\apimaptechnique}), and updating all existing code that used the old library’s API to now use the new library’s API while ensuring no change in the software’s behavior (\textit{\transform}).
This is a time-consuming and error-prone task that developers often dread \cite{kula2018developers}, especially in large codebases with pervasive use of the original library. 
Therefore, migration techniques that automate this entire process can save developers time and effort.

With the exception of a few techniques~\cite{ni2021soar}, most of the existing library migration literature targets Java libraries \cite{teyton2013automatic, alrubaye2018automating, alrubaye2019use, miningAnalogicalAPIs, zhang2020deep}.
Regardless of the supported language, most of these techniques stop at the API mapping~ stage~\cite{teyton2013automatic, alrubaye2018automating, alrubaye2019use, miningAnalogicalAPIs, zhang2020deep} and/or implicitly assume that function calls from the source library map to (one or more) function calls in the target library~\cite{teyton2013automatic, alrubaye2018automating, alrubaye2019use, miningAnalogicalAPIs, zhang2020deep,ni2021soar}.
When implementing \transform, researchers may also leverage certain aspects of the target library domain to implement their technique, e.g., the expressiveness of error messages in the data science domain~\cite{ni2021soar}.
However, designing general library migration support techniques requires a deep and systematic understanding of the required \codechange{s}.

\autoref{fig:f9f083ec} presents a Python code snippet migrated from the web application framework \lib{bottle} (left side) to an analogous library, \lib{flask} (right side).
An API mapping technique may suggest replacing the \texttt{post()} function with the \texttt{route()} function.
While helpful, the information is incomplete, because successfully transforming the shown client code to use the new library requires adding the \texttt{methods} argument with the value \texttt{[\textquotesingle POST\textquotesingle]} to the \texttt{route()} call.

Overall, there is no systematic knowledge of all types of API mappings and code changes for successful client code transformation from a source to a target library.
For example, some analogous libraries may have similar APIs, requiring minimal code modifications during migration.
Conversely, others may have considerable differences, requiring extensive modifications.
This raises several pertinent questions for the development of library migration tools and techniques.
\textit{What are the common types of code changes that developers typically perform to migrate from one library to another?}
\textit{Does a migration typically involve a single kind of code change or multiple related code changes?}
\textit{How difficult would it be to automate these changes?}
Overall, \textit{What are the characteristics of the code changes during library migrations?}

To the best of our knowledge, while there is a lot of literature on library migration, no study systematically characterizes \codechanges regardless of the targeted language.
Given the Python's rising popularity and its expanding library set \cite{pypi}, we address this gap by conducting an empirical study to understand different types of code changes that happen during Python library migration in open-source repositories.
We analyze real-world Python library migrations using two existing datasets, \migbench \cite{migbench} and \salm \cite{salm}, creating a new dataset \migbenchTwo.
While these datasets contain identified migration commits, they do not (identify or) analyze the involved code changes.
In this work, we identify migration-related code changes from the migrations in \migbenchTwo, label these code changes, and then build a taxonomy of code changes, \taxonomy. 
The resulting labeled migrations allow us to perform the empirical study and investigate the above questions; the answers to these questions can aid tool builders and researchers in developing Python library migration tools and techniques.

Our empirical study is based on \migbenchTwo, which includes \VTwoCCCount \codechanges in \VTwoMigCount Python library migrations from \VTwoRepoCount client repositories spanning \VTwoLPCount library pairs from \VTwoDomainCount domains.
We find that \HasNonFunctionLPPercent of library pairs have API mappings that involve program elements other than functions (e.g., \at or \de).
We also find that a program element from the source library (e.g., a function call) may be replaced by a different type of program element from the target library (e.g., an attribute access).
These results imply that some of the assumptions that current API mapping or client transformation tools make (e.g., that a function is always replaced by another function~\cite{teyton2013automatic, alrubaye2018automating,alrubaye2020learning, ni2021soar}) may not always hold in all migration scenarios.
Regardless of the type of program elements in an API mapping, \TotalElementNameChangePercent of API mappings involve program elements with \textit{different} names, and \FunctionCallFunctionCallArgumentTransformationPercent of \fc to \fc API mappings involve some form of argument transformation.
Thus, it is essential for tools to support a variety of types of code changes.
We also find that, on average (median), a migration includes \MigLOCMedian lines of code, \NumAPIsMedian API instances, and \NumChangesMedian code changes, and a developer who has to manually migrate client code needs to learn \UniqueAPIsMedian APIs and \UniqueMappingsMedian API mappings.
Thus, on one hand, the good news is that an automated tool may be able to automate the majority of migrations, given their simplicity and repetitiveness.
On the other hand, such automation requires handling several types of API mappings and argument transformations in certain more complex migrations.

To summarize, this paper makes the following contributions:
\begin{enumerate}[]
    \item We create a benchmark of \VTwoMigCount manually verified Python library migrations, \migbenchTwo, by expanding the original \migbench \cite{migbench} dataset with verified migrations from \salm \cite{salm}. 
    \item We manually label \VTwoCCCount \codechanges in the \migbenchTwo dataset to build \taxonomy, a taxonomy of Python library migrations.
    \item We conduct an empirical study on \migbenchTwo to characterize the nature of Python library migrations in terms of the required \codechanges and migration effort. 
\end{enumerate}

Researchers can use \migbenchTwo to evaluate their library migration tools and techniques and use \taxonomy for accurately labeling supported library migration types.
For example, one technique may only work for \transform that involves function mapping with no changes in arguments while another may be able to additionally handle argument changes and modifications to attribute access.
By clearly stating supported operations, techniques can be compared systematically and fairly.
Our findings, along with the discussed implications, provide insights into the characteristics of Python library migrations, highlighting the required migration support.
All our data and analysis scripts are available on our online artifact page \artifactURL.

\section{Background and Terminology}
\label{sec:background}

\subsection{Terminology and Notation}

\textit{Library migration} entails replacing one library with another in a software application, with the old being the \textit{source library} and the new being the \textit{target library}. Similarly, \textit{source API} and \textit{target API} refer to the APIs of these respective libraries. A library pair is termed \textit{analogous} if the two libraries provide similar functionality, allowing migration between them.

A commit in which migration occurs is a \textit{migration commit}. While migrations can span multiple commits \cite{alrubaye2020learning}, the datasets used in this study (see Section \ref{subsec:datasets}) consider only single-commit migrations. Therefore, for simplicity, this paper equates a migration with a migration commit. Nonetheless, a single commit can encompass multiple migrations between different library pairs.

During library migration, developers modify the client code to replace source library APIs with analogous target APIs. We refer to these modifications a \textit{\codechanges} or simply \textit{code changes}. A migration can require adjustments in multiple files and locations within a file. A single code change denotes a minimal API replacement that is indivisible into further meaningful replacements. For example, in \autoref{fig:f9f083ec},
we observe three distinct code changes.
The import on line 7 is replaced with the import in line 6, the function call \texttt{Bottle()} on line 68 is replaced by the function call to \texttt{Flask()} on line 67, and the function call \texttt{post()} on line 69 is replaced by a call to function \texttt{route()} on line 68.
We use the notation \ccLines{<removed-lines>}{<added-lines>} to specify a \ccs in a given diff; hence these three code changes are denoted by \ccLines{7}{6}, \ccLines{68}{67}, and \ccLines{69}{68}.
Since the same line number may appear in both the removed or added lines, we use \remLine{<line-number>} and \addLine{<line-number>} to denote removed and added lines, respectively.

By observing \ccs (along with reading the libraries' documentations), we can infer \textit{API mappings}, a term commonly used in the literature~\cite{ChenAPIMapping2021,NguyenAPIMapping16}, to indicate which API(s) from the target library perform the same functionality provided by the original API from the source library. From the changes in \autoref{fig:f9f083ec}, we can infer two API mappings \texttt{Bottle()} $\rightarrow$ \texttt{Flask()} and \texttt{post()} $\rightarrow$ \texttt{route()}. 
To successfully apply API mappings during \transform, developers may need to perform additional types of code changes. We refer to such additional code changes as \textit{properties}. For example, the code change that applied the API mapping from \texttt{Bottle()} to \texttt{Flask()} in \autoref{fig:f9f083ec} required the following properties: name change and argument addition. 

\subsection{Datasets}
\label{subsec:datasets}

We provide the background details of the two datasets we use: SALM \cite{salm} and \migbench \cite{migbench}.

\paragraph{SALM}
Gu \etal~\cite{salm} analyzed self-admitted library migrations (SALM) in Java, JavaScript and Python repositories to understand the nature and frequency of \libmigs.
Their contribution includes a dataset of self-admitted Python \libmigs which we refer to as \textit{SALM} in this paper.
Gu \etal use libraries.io \cite{libraries.io} to retrieve 10,147 popular Python libraries, and GHTorrent \cite{ghtorrent} to retrieve 121,381 client repositories for these libraries.
To identify self-admitted migrations, they use NLP-based heuristics to analyze the commit messages to identify commits that explicitly mention a migration between two libraries. 
They further filter out migrations between infrequent library pairs and
finally include 5,805 library migrations between 640 library pairs in 5,061 commits.

\paragraph{\migbench}
In our previous work~\cite{migbench}, we built \migbench, a benchmark of Python library migrations and locations of \ccs.
We used SEART \cite{dabic2021sampling} to retrieve a list of 195,075 non-forked Python repositories with 10 or more stars.
We first used various automated filtering to discard unlikely migrations, and then manually verified 1,244 migrations between library pairs that frequently appear in the migrations, resulting in \VOneMigCount migrations with code changes between \VOneLPCount analogous library pairs.

\section{Building \taxonomy}
\label{sec:pymigtax}

\newcommand{\labCell}{
    \multirow{3}{*}{\makecell{Labeled\\data}}
}

\newcommand{\fdCell}{
    \multirow{3}{*}{\makecell{Full data \\after filtering}}
}

\def\ROnePEKappa{0.43}
\def\ROneCardKappa{0.52}
\def\ROnePropAlpha{0.59}

\def\RTwoPEKappa{0.93}  
\def\RTwoCardKappa{0.94}
\def\RTwoPropAlpha{0.48}

\def\RThreePEKappa{0.97} 
\def\RThreeCardKappa{0.80}
\def\RThreePropAlpha{0.81}

\begin{table*}[t]
    \centering
    \caption{Datasets used in the study}
    \label{tab:dataset}
    \resizebox{.9\textwidth}{!}{
    \begin{tabular}{@{}llrrrrrrrr@{}}
    \toprule
    \textbf{Round} & \textbf{Dataset}   & \textbf{Migrations}   & \textbf{Repos}         & \textbf{Lib Pairs}   & \textbf{Libs}        & \textbf{Domains}          & $\kappa$ PE & $\kappa$ Cardinality & $\alpha$ Props\\ \midrule \midrule
    \fdCell        & \migbench          & \VOneMigCount         & \VOneRepoCount         & \VOneLPCount         & \VOneLibCount         & \VOneDomainCount         & & &\\
                   & \salm              & \SALMTotalMigCount    & \SALMTotalRepoCount    & \SALMTotalLPCount    & \SALMTotalLibCount    & \SALMTotalDomainCount    & & &\\ \cmidrule{2-7}
                   & All                & \GrandTotalMigCount   & \GrandTotalRepoCount   & \GrandTotalLPCount   & \GrandTotalLibCount   & \GrandTotalDomainCount   & & &\\ \midrule          
    Initial        & \migbench          & \VOneMigCount         & \VOneRepoCount         & \VOneLPCount         & \VOneLibCount         & \VOneDomainCount         & & &\\
    1              & \salm              & \RoundOneMigCount     & \RoundOneRepoCount     & \RoundOneLPCount     & \RoundOneLibCount     & \RoundOneDomainCount     & \ROnePEKappa   & \ROneCardKappa   & \ROnePropAlpha   \\
    2              & \salm              & \RoundTwoMigCount     & \RoundTwoRepoCount     & \RoundTwoLPCount     & \RoundTwoLibCount     & \RoundTwoDomainCount     & \RTwoPEKappa   & \RTwoCardKappa   & \RTwoPropAlpha   \\
    3              & \salm              & \RoundThreeMigCount   & \RoundThreeRepoCount   & \RoundThreeLPCount   & \RoundThreeLibCount   & \RoundThreeDomainCount   & \RThreePEKappa & \RThreeCardKappa & \RThreePropAlpha \\
    4              & \salm              & \RoundFourMigCount    & \RoundFourRepoCount    & \RoundFourLPCount    & \RoundFourLibCount    & \RoundFourDomainCount    & & &\\ \midrule
    \labCell       & \migbench          & \VOneMigCount         & \VOneRepoCount         & \VOneLPCount         & \VOneLibCount         & \VOneDomainCount         & & &\\
                   & \salm              & \SALMLabeledMigCount  & \SALMLabeledRepoCount  & \SALMLabeledLPCount  & \SALMLabeledLibCount  & \SALMLabeledDomainCount  & & &\\ \cmidrule{2-7}
                   & All                & \TotalLabeledMigCount & \TotalLabeledRepoCount & \TotalLabeledLPCount & \TotalLabeledLibCount & \TotalLabeledDomainCount & & &\\ \midrule
    Having code change & \migbenchTwo   & \VTwoMigCount         & \VTwoRepoCount         & \VTwoLPCount         & \VTwoLibCount         & \VTwoDomainCount         & & &\\ \midrule \bottomrule
    \end{tabular}%
}
\end{table*}

For our empirical study, we need a unified way to describe \codechanges.
Thus, as our first step, we use the identified migrations from the two datasets mentioned in Section~\ref{subsec:datasets}, \migbench and \salm, to derive a taxonomy of \ccs, referred to as \taxonomy.
We will now describe the steps undertaken to construct \taxonomy.

\subsection{Data curation}
\label{subsec:data-curation}

In our previous work, we manually validated each migration that is part of \migbench, enabling us to use it as is in this work.
However, \salm is an automatically constructed data set, which relies on heuristics to detect self-admitted migrations from commit messages.
Thus, we first verify the correctness of the data \salm provides, and apply the following filtering and preprocessing steps.

We remove 4 duplicate migrations found in \salm and focus on third-party libraries (the ones available in PyPI \cite{pypi}). There are 35 libraries in \salm not on PyPI, either due to misspellings or being system libraries, affecting 155 library pairs and resulting in the exclusion of 1,818 migrations. We observe some migrations in \salm between seemingly non-analogous library pairs, possibly due to its automated construction. An example includes migrations between \lib{flask} (a web application framework) and \lib{click} (CLI library). We identify 253 non-analogous library pairs in \salm using OpenAI's GPT-4 API \cite{gpt4} (prompt is available on the artifacts page) and exclude them and their corresponding migrations.
This results in \SALMTotalMigCount self-admitted \libmigs which we use.
The first three rows of \autoref{tab:dataset} summarize the two datasets after the above filtering. Given that some migrations overlap in both datasets, the \textit{All} row does not directly sum the preceding two rows.

\subsection{Building the initial version of \taxonomy}
\label{subsec:build-initial-taxonomy}
We build an initial version of \taxonomy based on the data available in \migbench because it specifies the line numbers of code changes for each recorded migration, unlike \salm.
For each migration in \migbench, 
we manually analyze the modified Python files in the corresponding commit to understand and label the types of code changes, following an iterative open-coding \cite{qualitativeMethods} process as follows.
The first author initially manually reviews code changes of 32 randomly selected migrations,
recording the source and target APIs and how the replacement between them is happening in the \codechanges (e.g., \textit{replace a function} and \textit{add an argument}). 
Based on these notes, the first author creates a draft of the taxonomy that they discuss with two other authors, looking at the corresponding examples and
discuss the structure and labels to use in the taxonomy.
Based on the discussion, the first author revisits and refines the draft taxonomy, applying the new insights to label additional data before another round of discussion.
This process continues until the three authors agree that the taxonomy accurately captures the observed \ccs available in \migbench. At the end of this process, we identify the following three dimensions for labeling \ccs: 

\begin{figure}[th]
    \centering
    \includegraphics[width=\diffWidth\columnwidth]{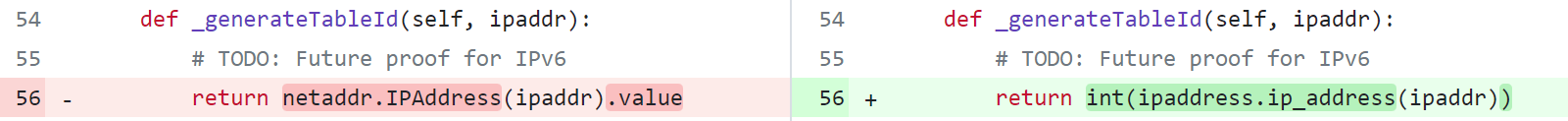}
    \includegraphics[width=\diffWidth\columnwidth]{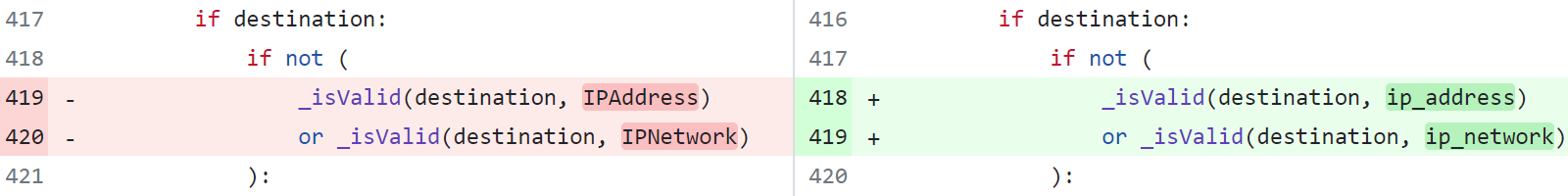}
    \caption{\migDesc{netaddr}{ipaddress}{ovirt/vdsm}{6eef802}}
    \label{fig:6eef802}
    \Description{}
\end{figure}

\paragraph*{Program elements} The program element are types of the APIs involved in the \cc, such as \fc and \at.
We further separate source and target program elements to indicate the types of source and target APIs, respectively.
For example, in the change \ccLines{56}{56} in \autoref{fig:6eef802}, the \fc \texttt{IPAddress} and the \at \texttt{value} from the source library are replaced with the \texttt{ip\_address} \fc from the target library. Here, the source program elements are \fc and \at, with the target program element being \fc.

\paragraph*{Cardinality} The numerical relationship between removed and added APIs in a \cc. 
We use \textit{many} to describe any quantity greater than one, as also used in the literature~\cite{alrubaye2019use}.
In the change \ccLines{56}{56} from \autoref{fig:6eef802}, two APIs are replaced with one, therefore, the cardinality is \mo.

\paragraph*{Properties} Additional specific details that characterize the \cc. 
Such properties emerged from our open coding process.
For example, in \autoref{fig:6eef802}, the new code on Line 56 requires a cast to an integer (using \texttt{int()}), which we label as \textit{\outTrans}.

We use the notation \ccDescLong{source program elements}{target program elements}{properties}{cardinality} to label a code change using \taxonomy, often omitting the cardinality.
For example, the change in line \ccLines{56}{56} in \autoref{fig:6eef802} can be denoted as \ccDesc{\fc, \at}{\fc}{\outTransShort}.
For brevity, we do not show the initial version of \taxonomy in the paper, but include it in our artifact.

\subsection{Extending \taxonomy}
While \migbench contains manually verified data, its small size poses a risk of constraining our taxonomy, potentially overfitting it solely to the migrations observed in this dataset.
To improve the generalizability of the taxonomy, we use the initial version of \taxonomy to label the code changes of a statistically representative sample of migrations in \salm.
We follow an iterative process (outlined below) where we allow modifications to \taxonomy based on the new data and then finalize the taxonomy when we reach saturation.

\subsubsection{Automatic Identification of Migration-related Changed Lines:}
\label{subsec:code-change-identification}
\salm does not include the location of the \codechanges and
might include commits with no actual changes, relying solely on self-admitted migration in commit messages.
Thus, we first automatically identify lines that may potentially have \ccs, focusing on those related to the source or target libraries in the self-admitted migration.
However, this requires detecting library usage in the modified files.
Although migrations in \salm are labeled with source and target library names, these names do not always match the import names; for instance, \textit{pyyaml} uses the import name \texttt{yaml}.
We use OpenAI's GPT-4 API \cite{gpt4} to retrieve the mappings between library names and their corresponding top-level import names (prompt is available on our artifacts page).
Given the import names, we can now identify usages of the libraries in the changed code. 
Specifically, we use the \texttt{ast} module to find usages of the libraries.
We then identify candidate migration lines as deleted lines where the source library was used or added lines where the target library is used.
We find a total of \CandidateSalmMigs of the \SALMTotalMigCount migrations in \salm with at least one file having candidate migration lines.
We do not find any \ccs in the remaining migrations. 

\subsubsection{Iterative Labeling Process} 
We find that the \CandidateSalmMigs migrations in \salm span \SALMTotalDomainCount library domains, as labeled by the original \salm authors.
Thus, when sampling the subset of migrations to manually analyze, we keep the following goals in mind: (1) the selected migrations are diverse across the different library domains to avoid biasing the types of code changes we observe to certain domains, 
(2) the sample size is statistically representative, and (3) the final selection is random. 

To satisfy these criteria, we perform a proportional stratified random selection~\cite{qualtricsSampling,KitchenhamSampling02, baltes2022sampling} as follows.
For statistical representativeness, we need to label \RequiredSALMSampleSize out of the total \CandidateSalmMigs migrations (5\% margin of error, 95\% confidence level), or 23.33\% of the migrations.
To ensure both diversity and randomness of the dataset, we perform this selection based on the migration count in each domain, selecting 23.33\% of migrations from each domain, and rounding up.
However, to ensure adequate data from all domains, we select a minimum of two migrations from each, unless only one is available.
Due to this approach, we finally select a total of \ActualSALMSampleSize migrations.
Our artifact depicts the distribution across application domains of the library pairs.

\begin{figure}[t!]
    \centering
    \includegraphics[width=.7\columnwidth]{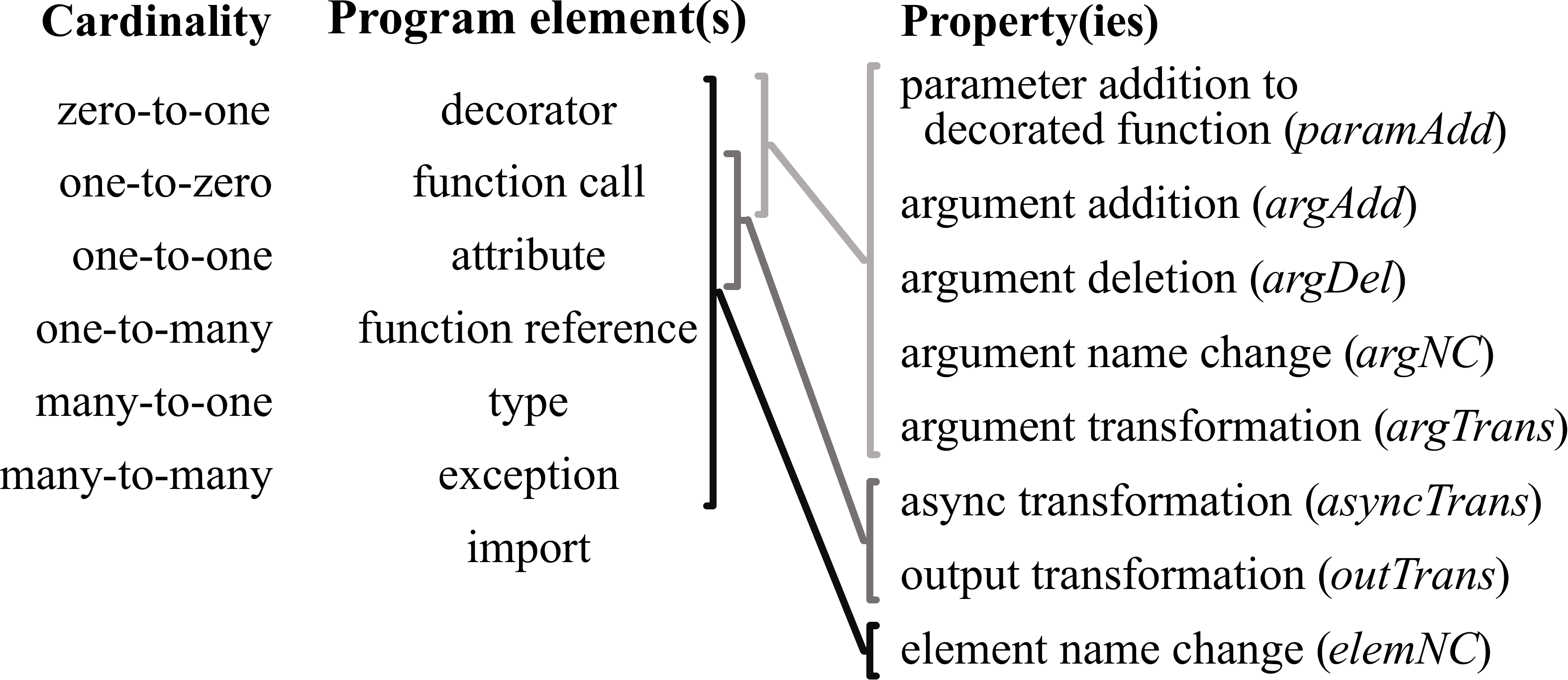}
    \caption{\taxonomy: a taxonomy of Python \codechanges}
    \label{fig:taxonomy}
    \Description{}
\end{figure}

We follow an iterative closed-coding approach \cite{qualitativeMethods} to label code changes in sampled migrations using the initial version of \taxonomy.
However, we also allow raters to identify new types of code changes.
In each round, we randomly select a number of migrations for labeling.
Each migration is independently labeled by two authors.
Disagreements are discussed and resolved after each round, involving all authors if needed.
This process continues until substantial inter-rater agreement is achieved, and no new taxonomy labels are identified.
Inter-rater agreement is measured using Cohen's kappa \cite{cohen1960coefficient} score for program elements and cardinality, aiming for a score of $\ge0.8$ for near-perfect agreement \cite{landis1977measurement}.
For properties, which can have multiple labels per code change, we use Krippendorff's alpha \cite{krippendorff2018content}, maintaining a customary threshold of $\ge0.8$.

The middle section of \autoref{tab:dataset} shows the labeled migrations per iteration and the corresponding kappa/alpha scores.
In the first round, we observe three new program elements and two new properties.
In the second round, we find three new properties but no new program elements.
In the third round, after labelling a total of 52 migrations with 371 \codechanges, we do not observe any new taxonomy items and also reach the target agreement for program elements, cardinality, and properties.
Subsequently, each of the remaining \RoundFourMigCount migrations, comprising \RoundFourCCCount related code changes, is labeled by a single author, with no new items added to the taxonomy.
The bottom section of \autoref{tab:dataset} summarizes the labeled data used to construct \taxonomy.

\subsection{A description of \taxonomy}

\begin{figure}[t!]
    \centering
    \includegraphics[width=\diffWidth\columnwidth]{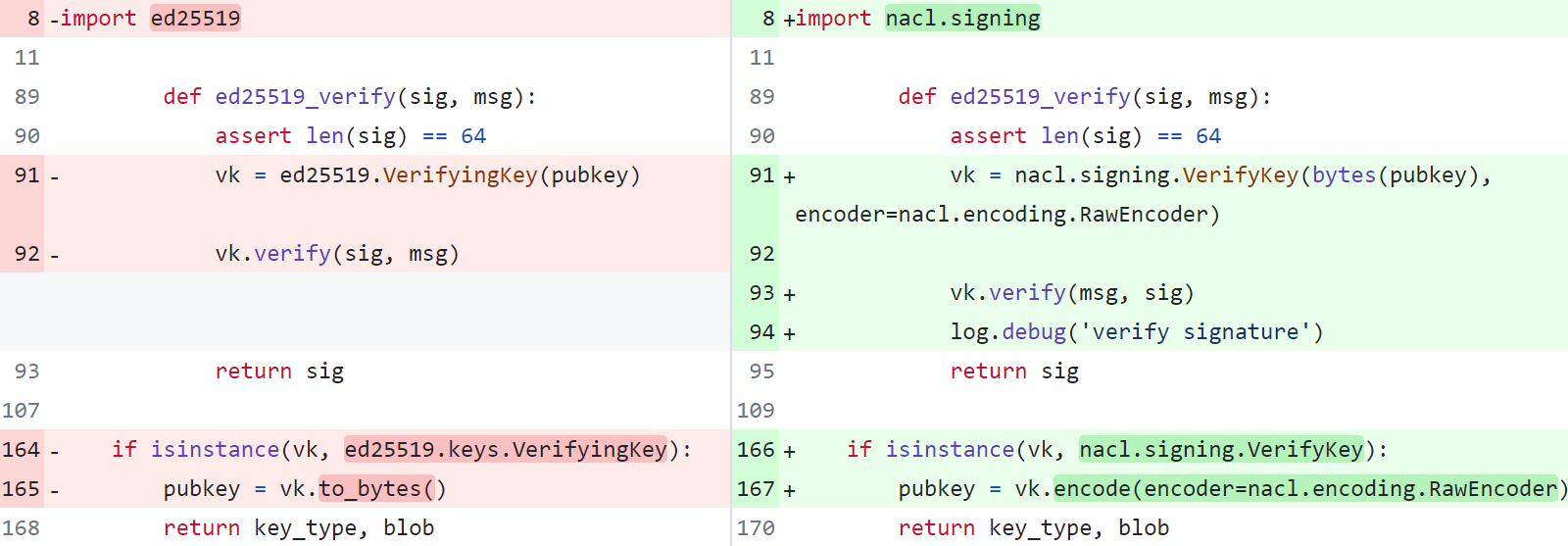}
    \caption{
        \migDesc{ed25519}{pynacl}{romanz/trezor-agent}{e1bbdb4}
    }
    \label{fig:e1bbdb4}
    \Description{}    
\end{figure}

\autoref{fig:taxonomy} illustrates the final version of \taxonomy, showing the three dimensions of a code change: cardinality (left), source/target program elements (middle), and properties (right).
These items can be combined in various ways to depict a given code change.
The connecting lines on the right show the observed properties for specific program elements, further detailed below.
The text in parentheses next to a property represents its short name, frequently used throughout the paper.

\subsubsection{Program elements}
Each \cc involves specific program elements, with seven distinct types observed in our data.
The first type, \textit{\fc}, includes \textit{calls} to functions, methods, constructors, etc., commonly referred to as callable\footnote{\url{https://docs.python.org/3/glossary.html\#term-callable}}.
\textit{\Ca}, on the other hand, occurs when a callable is only used as a reference, without calling it.
In \autoref{fig:6eef802}, note how the two uses of the \texttt{netaddr.IPAddress} function in lines \remLine{56} and \remLine{419} are labeled as \fc and \ca, respectively.

Similar to how \taxonomy distinguishes between \fcs and \ca{s}, it also distinguishes between a \textit{\ty} and a call to a constructor (the latter being a \fc).
In \autoref{fig:e1bbdb4}, \texttt{VerifyingKey} (\remLine{164}) is replaced with \texttt{VerifyKey} (\addLine{166}), with both the source and target program elements being \ty.
Libraries often use their own exception types, which also require migration.
Such program elements are labeled as \textit{\ex} in \taxonomy.

\begin{figure}
    \centering
    \includegraphics[width=\diffWidth\columnwidth]{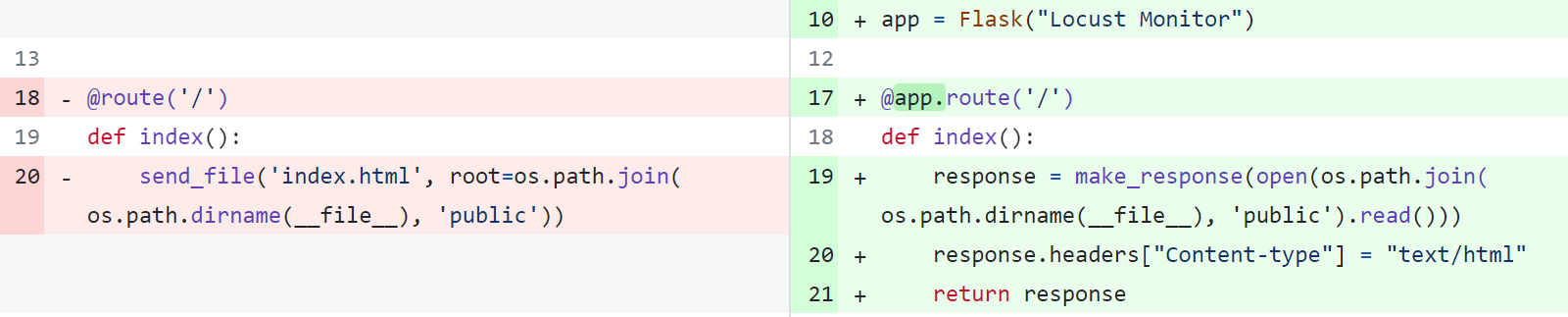}
    \caption{
        \migDesc{bottle}{flask}{heyman/locust}{4067b92}    
    }
    \label{fig:4067b92}
    \Description{}
\end{figure}

An \textit{\at} indicates access to an attribute of an object, as illustrated by the removal of access to the attribute \texttt{value} in line \remLine{56} of \autoref{fig:6eef802}.
A \textit{\de} denotes the use of a decorator, as shown in lines \remLine{18} and \addLine{17} of \autoref{fig:4067b92}.
During migration, developers typically replace the import statements of the source library with those of the target library, a change described by the \textit{\im} program element in \taxonomy, depicted in lines \ccLines{8}{8} of \autoref{fig:e1bbdb4}.
Given the stylistic variations in importing a library or its APIs, \im code changes are not labeled with cardinality or properties.

\subsubsection{Cardinality}
We find code changes having a total 6 different cardinalities, including \oo, \om, \mo, and \mm, which are frequently mentioned in the literature \cite{alrubaye2018automating,teyton2014study}.
Cases also exist where APIs are either only removed or added, indicating \oz or \zo cardinalities, respectively.
\autoref{fig:4067b92} showcases a migration between the web application frameworks \lib{bottle} and \lib{flask}, illustrating a \zo cardinality.
Here, the source library, \lib{bottle}, uses a static method (\texttt{route}) for routing, while the target library, \lib{flask}, uses an instance method (\texttt{Flask.route}), requiring the initialization of the \textit{Flask} object in line \addLine{10}.

\subsubsection{Properties}
\begin{figure}
    \centering
    \includegraphics[width=\diffWidth\columnwidth]{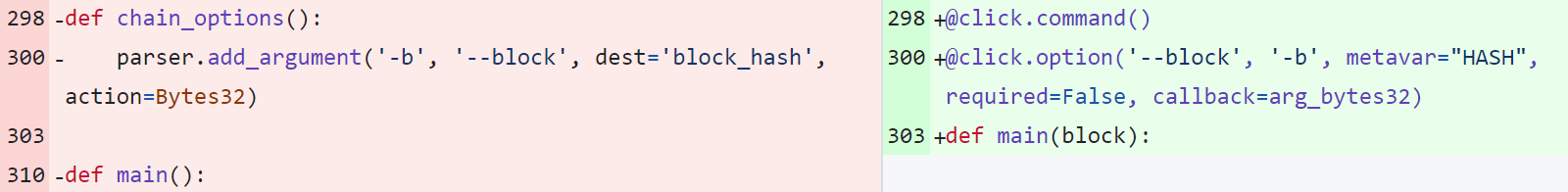}
    \caption{
        \migDesc{argparse}{click}{clearmatics/ion}{03fb3a3}.
    }
    \label{fig:03fb3a3}
    \Description{}
\end{figure}

Each \cc can additionally be characterized by properties. The property \textit{\enc} applies when the names of the equivalent source and target APIs differ, such as the replacement of the \fc \texttt{Bottle()} with the \fc \texttt{Flask()} in \autoref{fig:f9f083ec}, which involves an \enc.

Code change \ccLines{300}{300} in \autoref{fig:03fb3a3} exhibits several argument-related properties in the replacement of \texttt{add\_argument()} with the \de \texttt{@option()}.
\textit{\ArgNC} apply to instances where the source and target APIs have identical arguments, but their names differ, as seen with \texttt{action} and \texttt{callback}.
\textit{\ArgAdd} occurs when one or more arguments are added to the target \fc or \de during migration, exemplified by \texttt{metavar} and \texttt{required}.
Conversely, \textit{\argDel} involves the removal of an argument, such as \texttt{dest}.

Notice how the newly decorated function \texttt{main} in \autoref{fig:03fb3a3} now takes a new parameter (\texttt{block}) to replace the original command line arguments with the same names in \remLine{300}.
\taxonomy has the property \textit{\paramAddToDecorate} to describe this change.
Note that \paramAddToDecorate differs from \argAdd as the former affects a client code function, while the latter applies to the target API.

Sometimes, added and removed program elements take similar arguments, but argument changes are needed during migration.
\taxonomy uses the \textit{\argTrans} property to label such changes.
\autoref{fig:e1bbdb4} illustrates this with \texttt{VerifyingKey} and \texttt{VerifyKey} from source and target libraries, both requiring a key, but the latter expects it in byte format, necessitating an additional \texttt{bytes} call on line \addLine{91}.
Another example involves the \texttt{verify} methods in both libraries, taking identical arguments but with reversed positions.
\autoref{fig:4067b92} depicts a third instance, where \texttt{send\_file} and the corresponding \texttt{make\_response} function in the target library have differing requirements.

\taxonomy features the \textit{\outTrans} property to address cases where source and target APIs return data in differing formats, thus requiring transformation during migration.
In \autoref{fig:6eef802}, the source API provides a \texttt{value} attribute to access the returned IP address as an integer, whereas the target library recommends using the built-in \texttt{int} function for integer representation, exemplifying an output transformation property through the \texttt{int} call.
The \textit{\asyncChange} applies when either source or target APIs use async/await keywords, as shown lines \ccLines{36}{38} in \autoref{fig:1d8923a}.

Note that some code changes may not have any properties.
For example, the \des replaced in \ccLines{18}{17} of \autoref{fig:4067b92} have identical names and signatures, a code change with no properties.

\subsection{\migbenchTwo}
\label{subsec:migbench2}
We used \migbench \cite{migbench} and a subset of \salm \cite{salm} to build \migbenchTwo. 
Similar to \migbench~\cite{migbench}, we store our labeled data in a YAML format, one YAML file for each migration.
While the original \migbench included only the locations of each code change, \migbenchTwo includes the program elements, cardinality, and the additional properties we label using \taxonomy.
Additionally, a code change entry includes the names of the source and target APIs removed and added, respectively, in this code change.
The ``Having code change'' row in \autoref{tab:dataset} shows the data included in \migbenchTwo.
The dataset and its documentation are available in our artifact.

\section{Empirical study}
\label{sec:results}
In our empirical study, we answer three main research questions.

\def\RQOne{What are the common types of API mappings that appear during migration between Python libraries?\xspace}
\def\SubRQProgramElements{What program elements are typically involved in API mappings?}
\def\RQCombination{What combinations of code changes are common in Python library migrations?}
\def\RQEffort{How much development effort is needed for Python library migrations?}

\begin{enumerate}[label=\textbf{RQ\arabic*},leftmargin=*]
    \item \label{rq:code-changes} \RQOne 
    \begin{enumerate}[label*=.\textbf{\arabic*},leftmargin=*]
        \item \label{subrq:program-elements} \SubRQProgramElements
        \item \label{subrq:properties} What properties are typically involved in the code transformations for different types of API mappings?
  	\end{enumerate}
    \item \label{rq:combination} \RQCombination
    \item \label{rq:effort} \RQEffort  
\end{enumerate}

\subsection{\ref{rq:code-changes}~\RQOne}
\begin{table*}[t!]
  \centering
  \caption{Distribution of types of API mappings in \migbenchTwo}
  \label{tab:taxonomy-dist}\includegraphics[width=\textwidth]{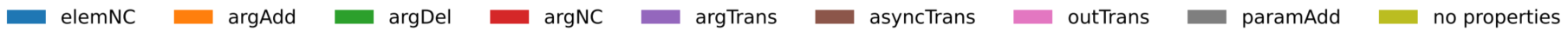}
  \resizebox{\textwidth}{!}{
    \begin{tabular}{llccccccccc}\hline
       &  & \multicolumn{9}{c}{\textbf{Target program elements}} \\ 
      \textbf{} & \textbf{} & \textbf{function call} & \textbf{attribute} & \textbf{decorator} & \textbf{function reference} & \textbf{type} & \textbf{exception} & \textbf{import} & \textbf{none} & \textbf{total} \\ \cmidrule{3-11}
      \multirow{18}{*}{\rotatebox{90}{\textbf{Source program elements}}} & \textbf{function call} & \makecell{\FunctionCallFunctionCallLPCount~(\FunctionCallFunctionCallLPPercent) \\ \includegraphics[width=1.3cm, height=0.5cm]{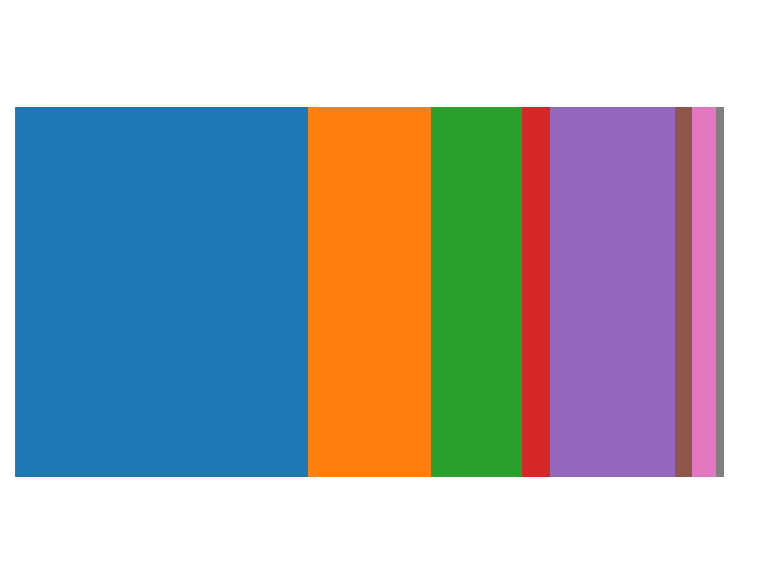}} & \makecell{\FunctionCallAttributeLPCount~(\FunctionCallAttributeLPPercent) \\ \includegraphics[width=1.3cm, height=0.5cm]{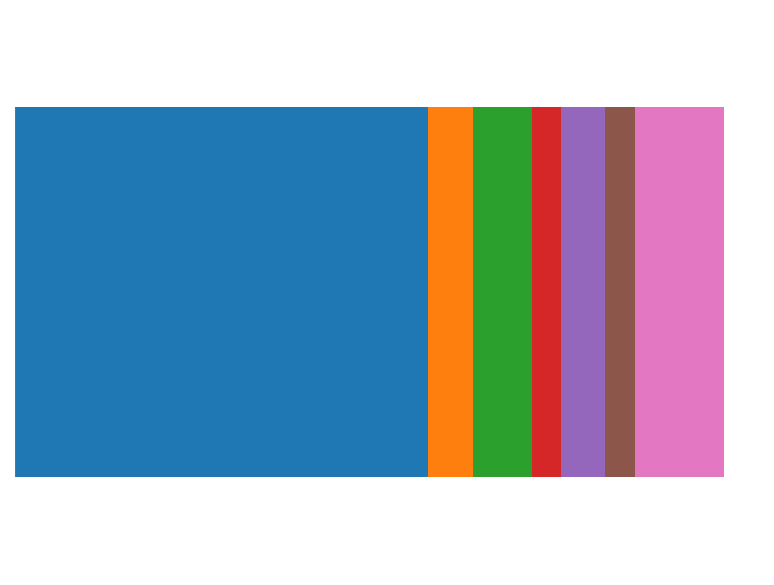}} & \makecell{\FunctionCallDecoratorLPCount~(\FunctionCallDecoratorLPPercent) \\ \includegraphics[width=1.3cm, height=0.5cm]{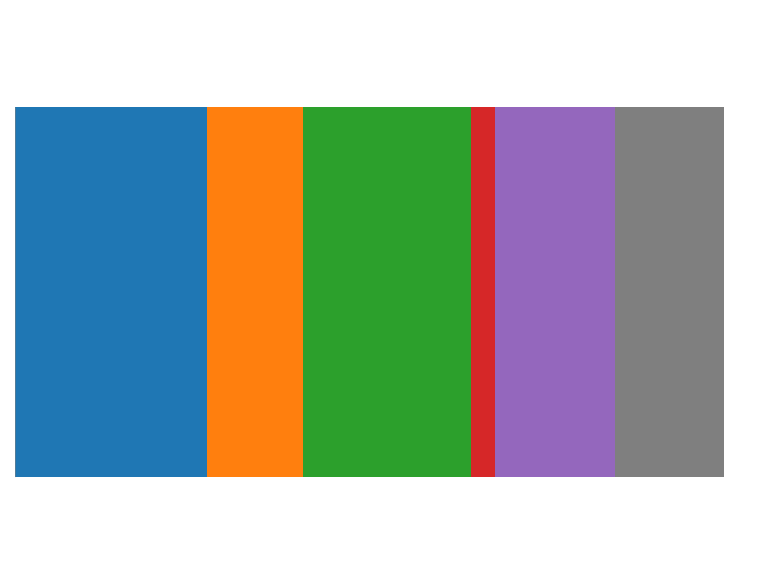}} & - & - & - & - & \FunctionCallNoneLPCount~(\FunctionCallNoneLPPercent) & \makecell{\FunctionCallTotalLPCount~(\FunctionCallTotalLPPercent) \\ \includegraphics[width=1.3cm, height=0.5cm]{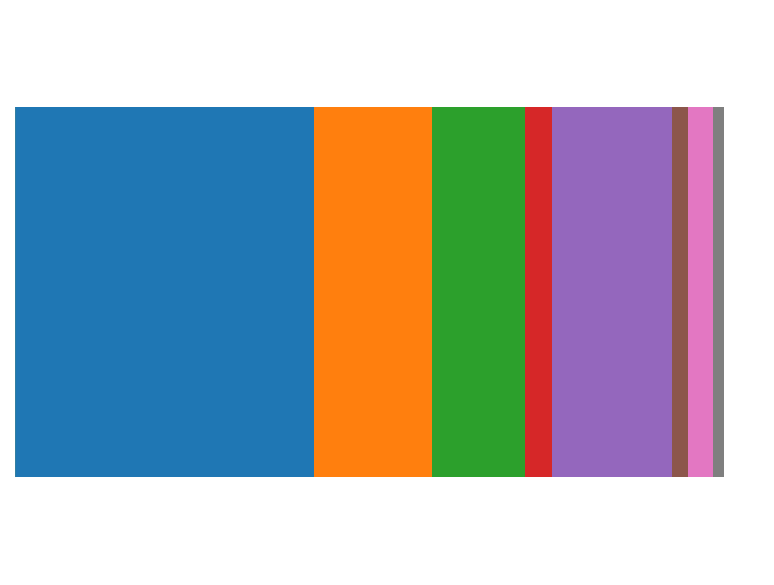}} \\ \cmidrule{3-11}
       & \textbf{attribute} & \makecell{\AttributeFunctionCallLPCount~(\AttributeFunctionCallLPPercent) \\ \includegraphics[width=1.3cm, height=0.5cm]{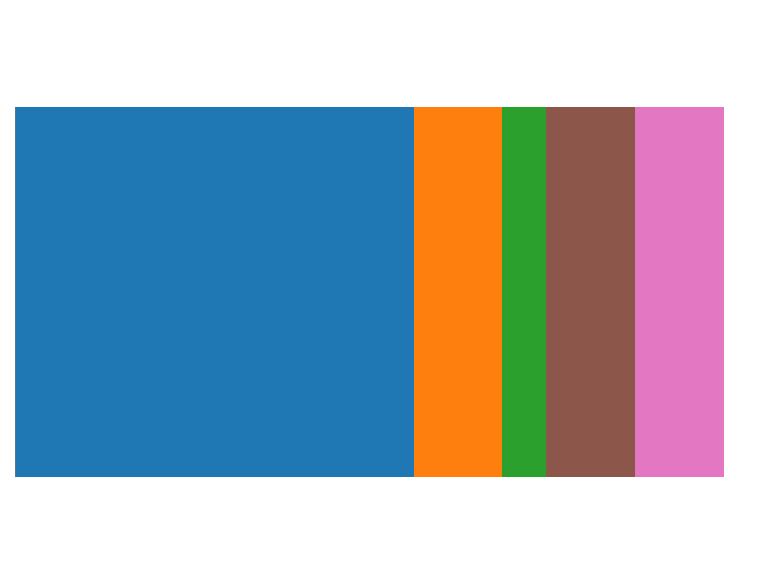}} & \makecell{\AttributeAttributeLPCount~(\AttributeAttributeLPPercent) \\ \includegraphics[width=1.3cm, height=0.5cm]{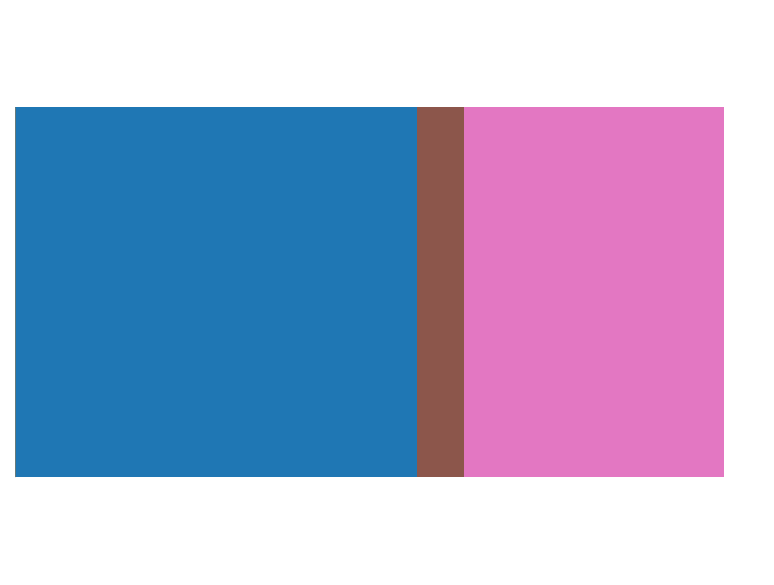}} & - & - & - & - & - & \AttributeNoneLPCount~(\AttributeNoneLPPercent) & \makecell{\AttributeTotalLPCount~(\AttributeTotalLPPercent) \\ \includegraphics[width=1.3cm, height=0.5cm]{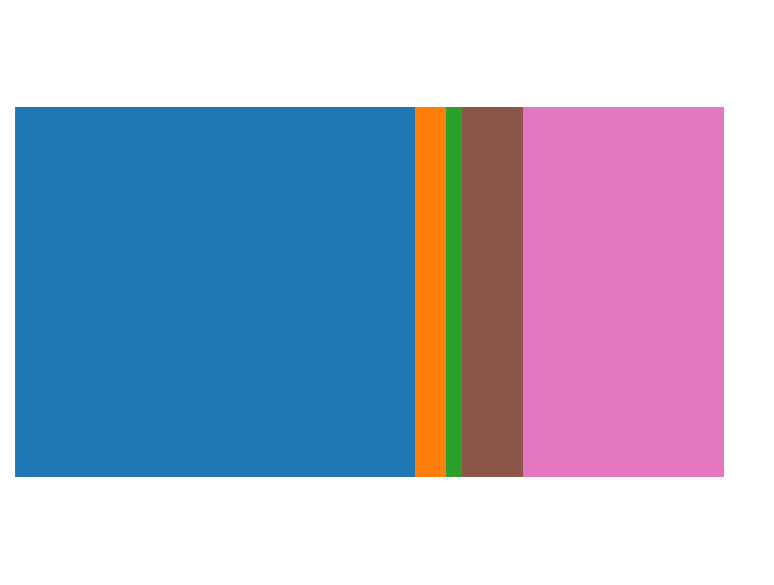}} \\ \cmidrule{3-11}
       & \textbf{decorator} & \makecell{\DecoratorFunctionCallLPCount~(\DecoratorFunctionCallLPPercent) \\ \includegraphics[width=1.3cm, height=0.5cm]{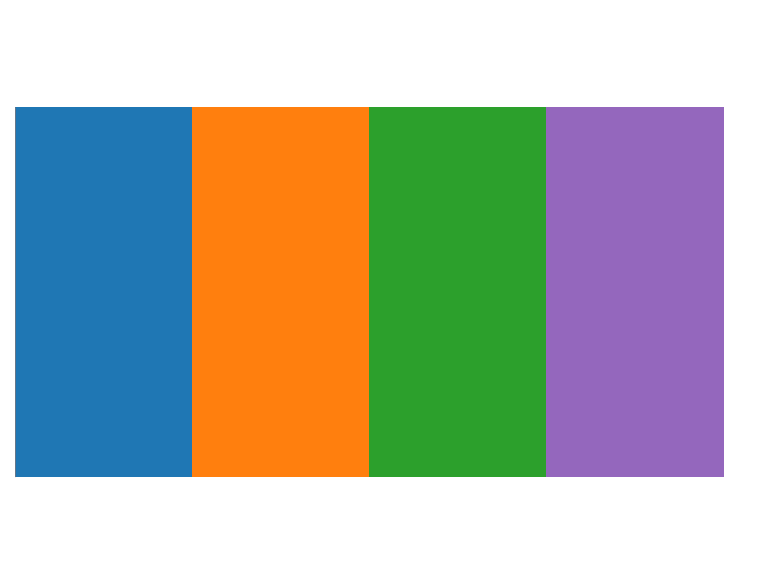}} & - & \makecell{\DecoratorDecoratorLPCount~(\DecoratorDecoratorLPPercent) \\ \includegraphics[width=1.3cm, height=0.5cm]{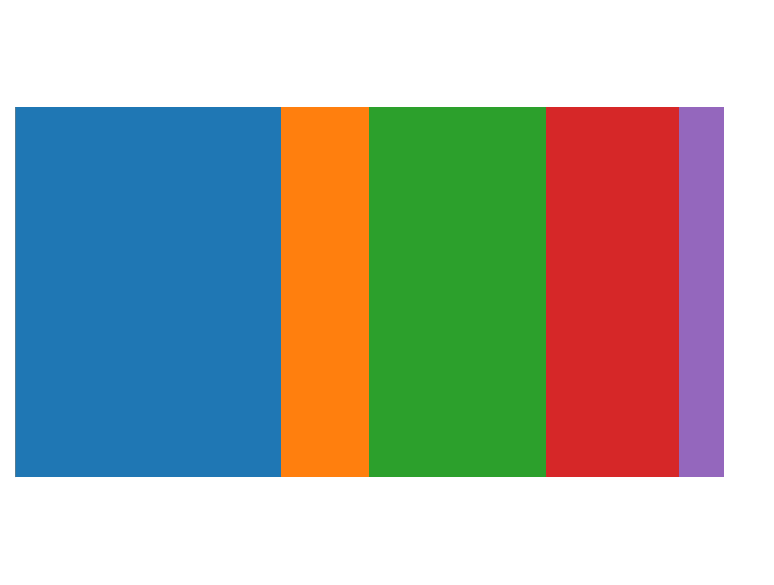}} & - & - & - & - & \DecoratorNoneLPCount~(\DecoratorNoneLPPercent) & \makecell{\DecoratorTotalLPCount~(\DecoratorTotalLPPercent) \\ \includegraphics[width=1.3cm, height=0.5cm]{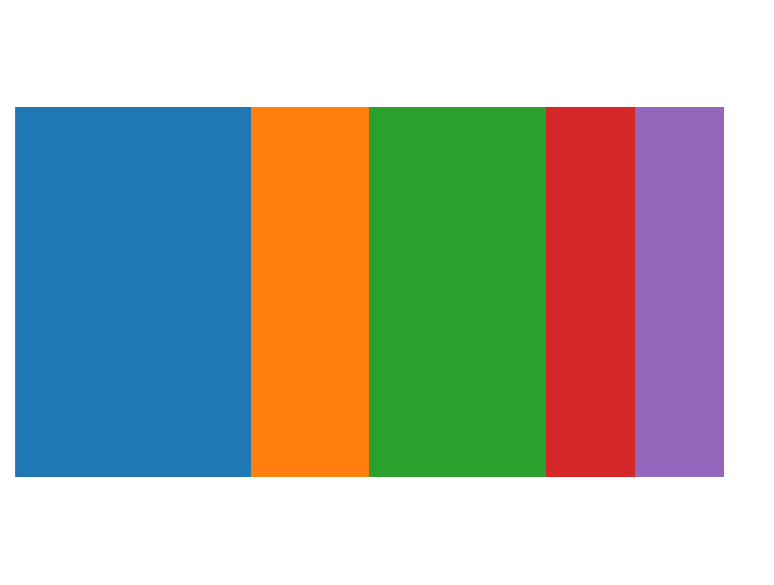}} \\ \cmidrule{3-11}
       & \textbf{function reference} & - & - & - & \makecell{\FunctionReferenceFunctionReferenceLPCount~(\FunctionReferenceFunctionReferenceLPPercent) \\ \includegraphics[width=1.3cm, height=0.5cm]{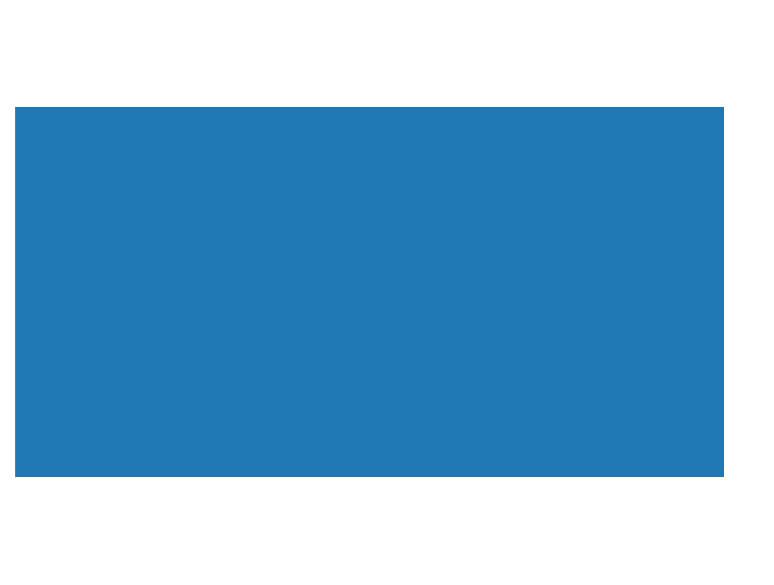}} & - & - & - & \FunctionReferenceNoneLPCount~(\FunctionReferenceNoneLPPercent) & \makecell{\FunctionReferenceTotalLPCount~(\FunctionReferenceTotalLPPercent) \\ \includegraphics[width=1.3cm, height=0.5cm]{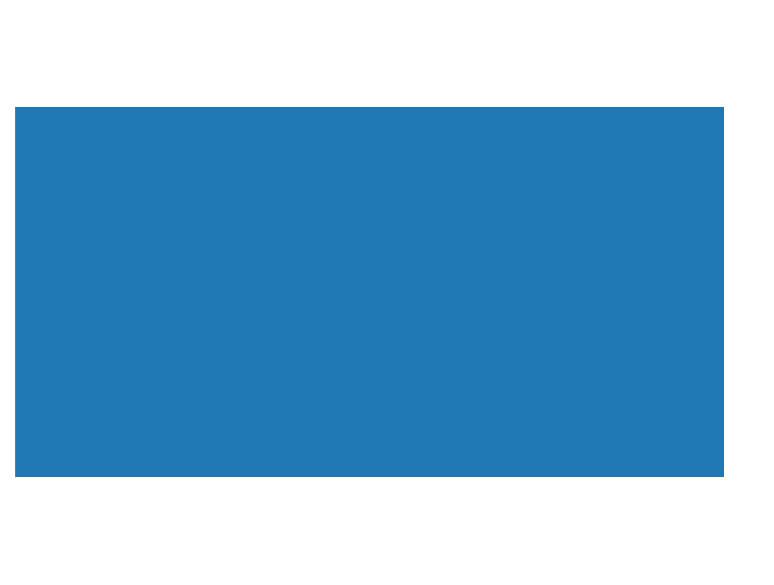}} \\ \cmidrule{3-11}
       & \textbf{type} & - & - & - & - & \makecell{\TypeTypeLPCount~(\TypeTypeLPPercent) \\ \includegraphics[width=1.3cm, height=0.5cm]{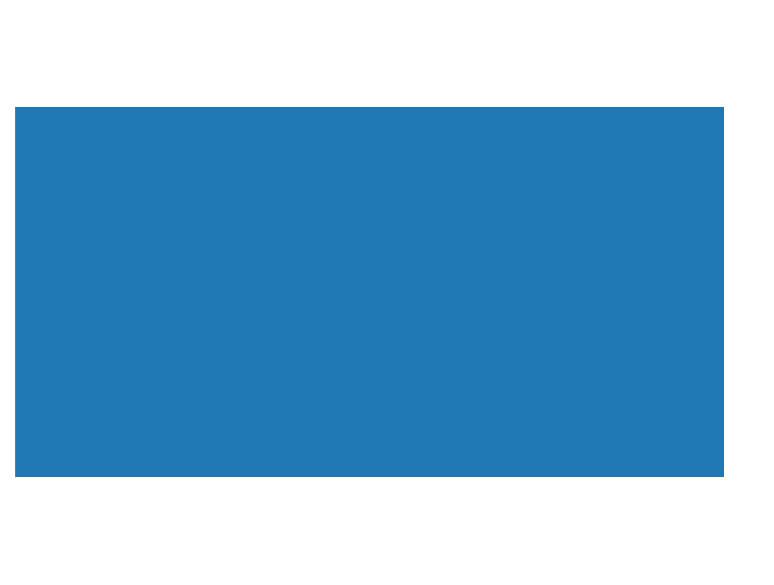}} & - & - & \TypeNoneLPCount~(\TypeNoneLPPercent) & \makecell{\TypeTotalLPCount~(\TypeTotalLPPercent) \\ \includegraphics[width=1.3cm, height=0.5cm]{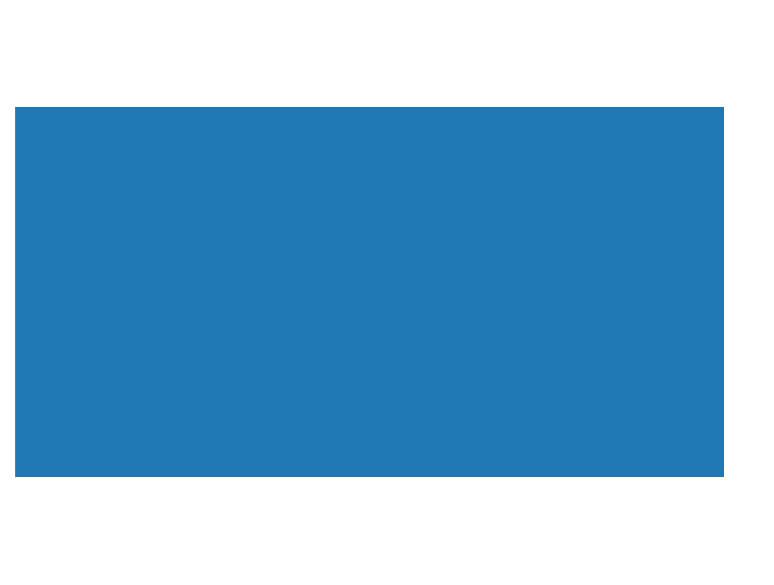}} \\ \cmidrule{3-11}
       & \textbf{exception} & - & - & - & - & - & \makecell{\ExceptionExceptionLPCount~(\ExceptionExceptionLPPercent) \\ \includegraphics[width=1.3cm, height=0.5cm]{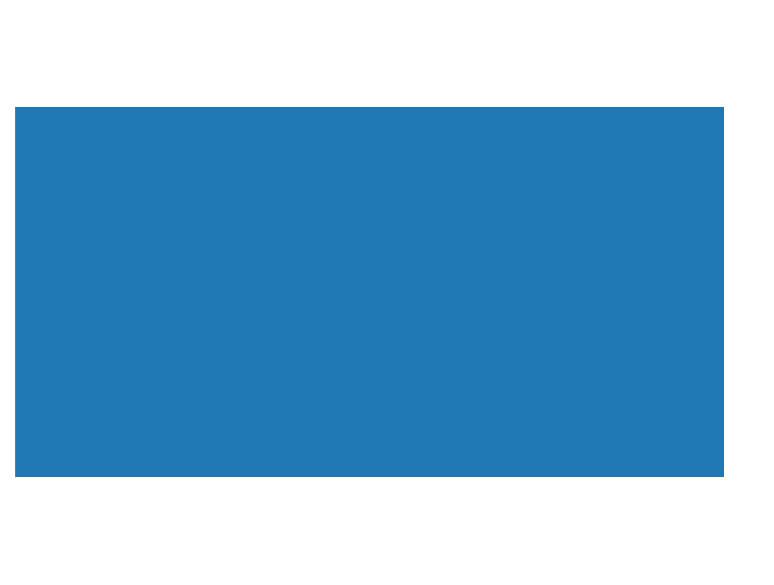}} & - & \ExceptionNoneLPCount~(\ExceptionNoneLPPercent) & \makecell{\ExceptionTotalLPCount~(\ExceptionTotalLPPercent) \\ \includegraphics[width=1.3cm, height=0.5cm]{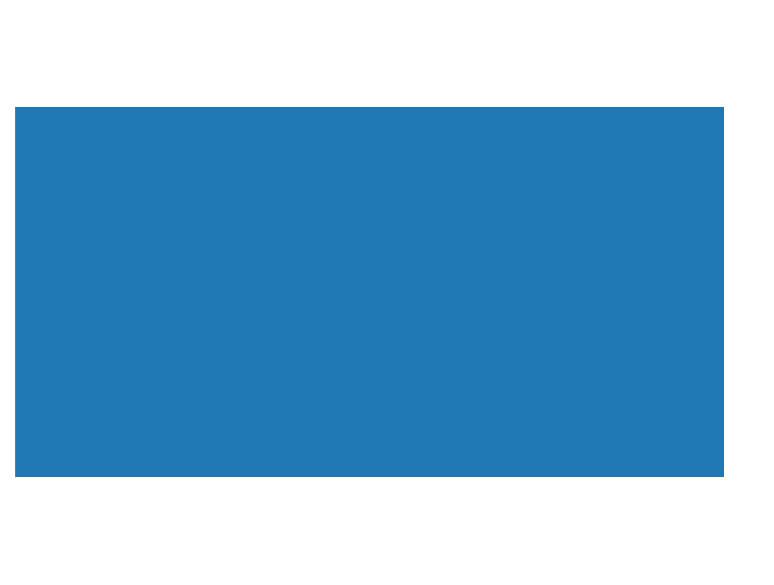}} \\ \cmidrule{3-11}
       & \textbf{import} & - & - & - & - & - & - & \ImportImportLPCount~(\ImportImportLPPercent) & \ImportNoneLPCount~(\ImportNoneLPPercent) & \ImportTotalLPCount~(\ImportTotalLPPercent) \\ \cmidrule{3-11}
       & \textbf{none} & \NoneFunctionCallLPCount~(\NoneFunctionCallLPPercent) & \NoneAttributeLPCount~(\NoneAttributeLPPercent) & \NoneDecoratorLPCount~(\NoneDecoratorLPPercent) & \NoneFunctionReferenceLPCount~(\NoneFunctionReferenceLPPercent) & \NoneTypeLPCount~(\NoneTypeLPPercent) & \NoneExceptionLPCount~(\NoneExceptionLPPercent) & \NoneImportLPCount~(\NoneImportLPPercent) & - & \NoneTotalLPCount~(\NoneTotalLPPercent) \\ \cmidrule{3-11}
       & \textbf{total} & \makecell{\TotalFunctionCallLPCount~(\TotalFunctionCallLPPercent) \\ \includegraphics[width=1.3cm, height=0.5cm]{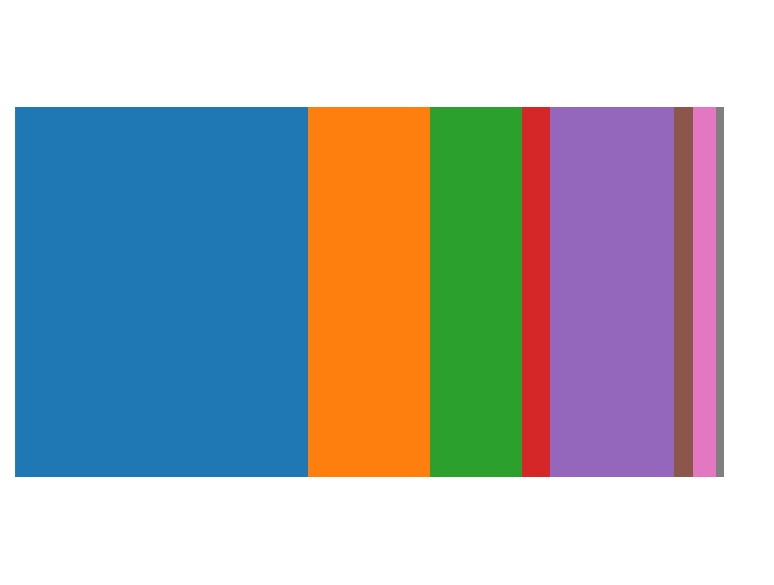}} & \makecell{\TotalAttributeLPCount~(\TotalAttributeLPPercent) \\ \includegraphics[width=1.3cm, height=0.5cm]{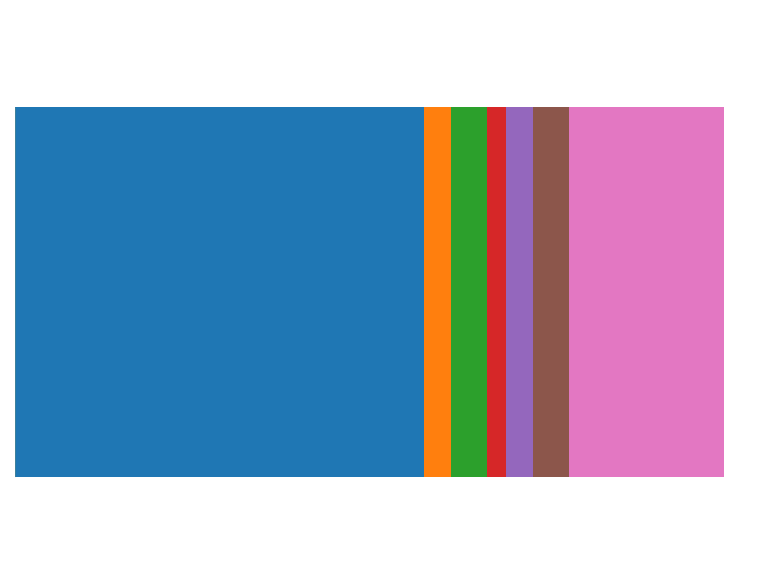}} & \makecell{\TotalDecoratorLPCount~(\TotalDecoratorLPPercent) \\ \includegraphics[width=1.3cm, height=0.5cm]{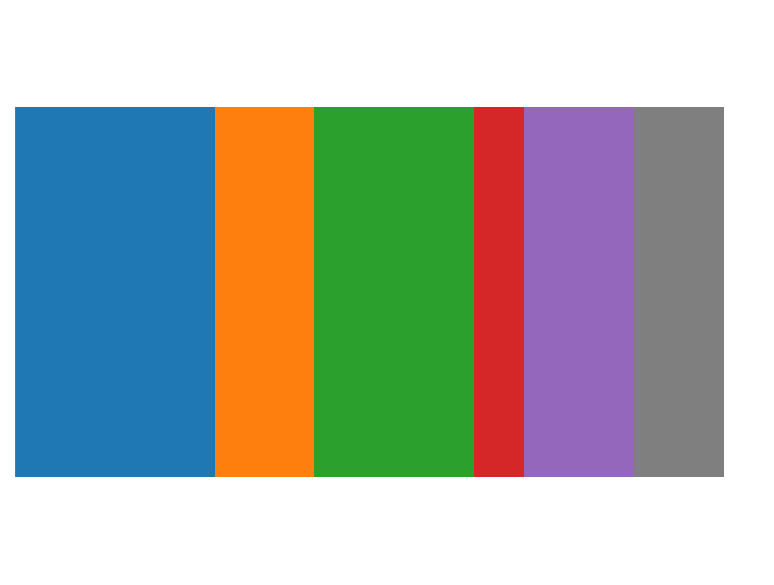}} & \makecell{\TotalFunctionReferenceLPCount~(\TotalFunctionReferenceLPPercent) \\ \includegraphics[width=1.3cm, height=0.5cm]{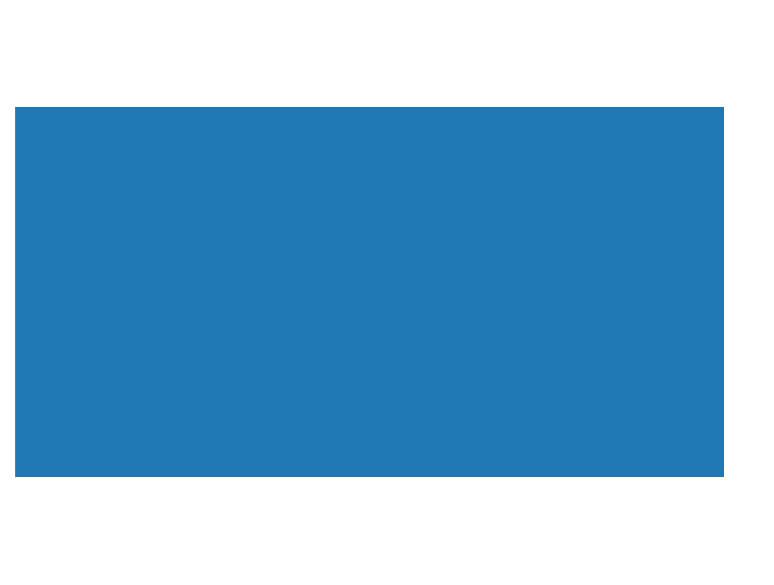}} & \makecell{\TotalTypeLPCount~(\TotalTypeLPPercent) \\ \includegraphics[width=1.3cm, height=0.5cm]{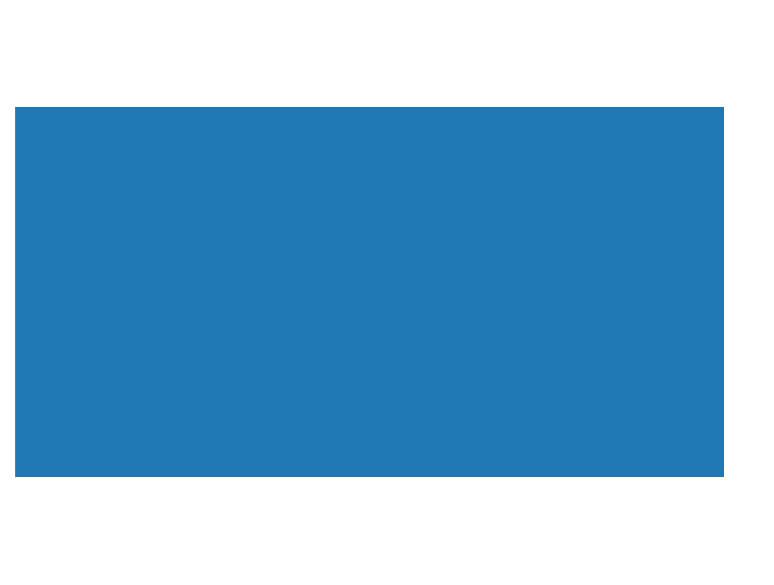}} & \makecell{\TotalExceptionLPCount~(\TotalExceptionLPPercent) \\ \includegraphics[width=1.3cm, height=0.5cm]{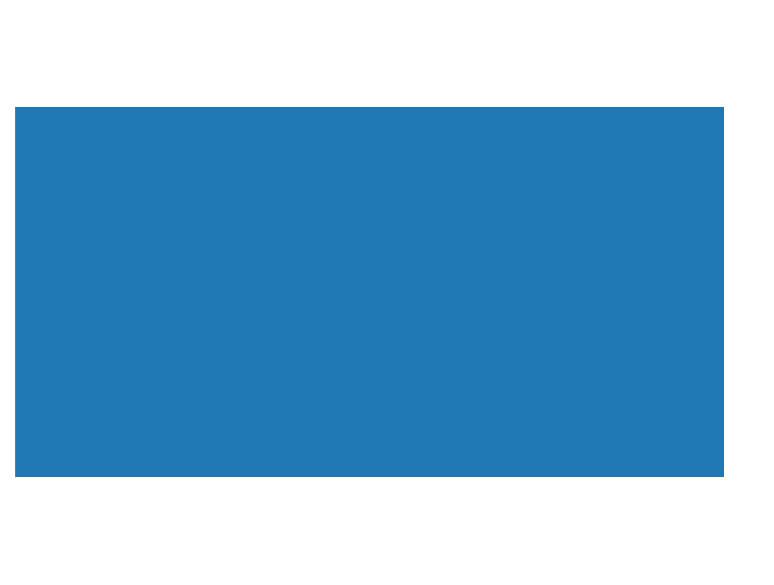}} & \TotalImportLPCount~(\TotalImportLPPercent) & \TotalNoneLPCount~(\TotalNoneLPPercent) & \makecell{\TotalLPCount~(\TotalLPPercent) \\ \includegraphics[width=1.3cm, height=0.5cm]{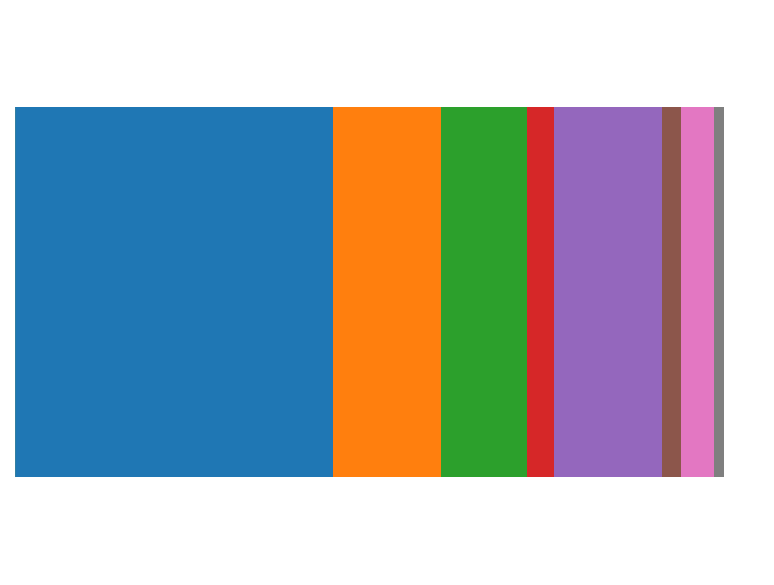}} \\ \hline
    \end{tabular}}
\end{table*}

\subsubsection{Motivation.}  Migrating from one library to another requires modifying the code from using the old to the new library.
While some analogous libraries have similar APIs, requiring minimal adjustments, others may necessitate extensive modifications due to differing APIs.
The data we labeled using \taxonomy helps us to understand typical \textit{API mappings} from source to target libraries.

\subsubsection{Methods.} 
We answer this RQ by analyzing the distribution of program elements and properties that appear in the API mappings we infer from the \cc in \migbenchTwo.

Consistent with existing literature \cite{teyton2013automatic,alrubaye2018automating,alrubaye2019migrationminer}, we derive API mappings from the code changes we observe.
For example, the code changes \ccLines{91}{91} and \ccLines{164}{166} in \autoref{fig:e1bbdb4} both replace \texttt{VerifyingKey} with \texttt{VerifyKey}.
Therefore, \texttt{VerifyingKey} $\rightarrow$ \texttt{VerifyKey} is an API mapping.
An API mapping may appear in several code changes, or even several migrations.
In that case, the corresponding code changes will all have the same source and target APIs; however, the properties involved in the code change may be different.
The specific properties for each code change depend on the underlying client code and do not change the API mapping.
Therefore, when studying API mappings in \ref{rq:code-changes}, we simply combine all observed properties for an API mapping.
Suppose \textit{foo()} $\rightarrow$ \textit{bar()} API mapping requires element name change in one migration and argument addition and element name change in another migration, then we would have a single set of properties \textit{\{element name change, argument addition\}} for this API mapping. 
Due to this difference in the nature of program elements and the corresponding differences in data analysis, we divide this RQ into two research questions for clarity.

In \ref{subrq:program-elements}, given a library pair, we aim to identify the types of source and target program elements that appear in the API mappings for that library pair.
\autoref{tab:taxonomy-dist} presents our findings for \ref{subrq:program-elements}, as the top number and percentage in each cell.
The columns and rows depict different types of program elements from \taxonomy.
The number at the top of each cell shows the number of \textit{library pairs} for which we find at least one API mapping between the respective program elements.
The percentage is then calculated from the total of \TotalLPCount unique library pairs in the dataset.
For example, the first cell in row two indicates that \AttributeFunctionCallLPCount (\AttributeFunctionCallLPPercent) library pairs have at least one API mapping where an \at is replaced with a \fc.
\textit{None} indicates no source/target elements, representing API additions/removals, and a dash (-) denotes combinations of elements that we did not observe.

In \ref{subrq:properties}, we explore the properties that occur during client code transformation.
Specifically, we are interested to know out of all observed API mappings between two types of program elements (regardless of which library pair they belong to), how many API mappings required a specific property during the corresponding client code transformations.
Thus, it is important to note that for properties, the unit of measurement is \textit{not} a library pair, but rather an API mapping.
We also use \autoref{tab:taxonomy-dist} to summarize the results, focusing on the stacked bar charts for properties.
The table's colored legend shows different properties, and each cell displays the distribution of these properties for API mappings between the respective program elements.
For example, the table reveals that \textit{all} API mappings that required changing a \ty from the source library to a \ty in the target library only exhibited \enc.

\subsubsection{\ref{subrq:program-elements} Findings (Program Elements in library pairs)}
We now discuss the top number in each cell from \autoref{tab:taxonomy-dist} related to program elements.
We observe that \im replacements are found in almost all library pairs (\ImportImportLPPercent), as libraries typically have distinct import names, prompting changes during migration.
However, it is interesting to note that some libraries, aiming for easy migration, maintain identical import names. We find \LibPairsWithSameImportNameCount such pairs in \migbenchTwo: \lib{PyCryptoDome} and \lib{PyCrypto}, both using the import name \texttt{Crypto}, and \lib{py-bcrypt} and \lib{bcrypt}, both using \texttt{bcrypt}.

Given that functions are the main type of API offered by libraries, it is not surprising to see that most library pairs (\FunctionCallTotalLPPercent) have API mappings involving \fcs as a source element, and \TotalFunctionCallLPPercent as a target element.
Interestingly, \fcs are not always replaced by other \fcs; they can also be replaced by \ats (\FunctionCallAttributeLPPercent of library pairs) and, less frequently, \des (\FunctionCallDecoratorLPPercent of library pairs).
The opposite is also true: \ats can be replaced with \fcs (\AttributeFunctionCallLPPercent) and \des replaced by \fcs (\DecoratorFunctionCallLPPercent). 

Replacements involving \ca, \ty, and \ex replacements are less common, found in only \FunctionReferenceTotalLPPercent, \TypeTotalLPPercent, and \ExceptionTotalLPPercent migrations respectively.
Function reference and \ty are only replaced with the same type of program element or are removed. While most migrations replace source APIs with target APIs, we also observe some cases of addition (\textit{none} column,  \TotalNoneLPPercent library pairs) and removal (\textit{none} row, \NoneTotalLPPercent library pairs).

\subsubsection{\ref{subrq:properties} Findings (Properties in API mappings)}
Turning to the properties in code transformations, the predominant blue in in the stacked bar charts of~\autoref{tab:taxonomy-dist} highlights frequent \enc. It suggests that the APIs from the source library do not necessarily map to APIs with the same name in the target library.
Nevertheless, \FunctionCallFunctionCallSameNamePercent of \fc to \fc API mappings  retained the same name in both libraries.

Argument-related properties only apply to \fc{s} and \de{s}, as other program elements do not take arguments.
We observe \argAdd (orange) in \FunctionCallFunctionCallArgumentAdditionPercent of \fc to \fc API mappings and \FunctionCallDecoratorArgumentAdditionPercent~of \fc to \des API mappings.
We also see that \argAdd occurs in \DecoratorFunctionCallArgumentAdditionPercent of \des to \fc API mappings and \DecoratorDecoratorArgumentAdditionPercent of \des to \des API mappings.
We find that \argDel (green) is less common than \argAdd (total \TotalArgumentDeletionPercent vs \TotalArgumentAdditionPercent).
Specifically, they are common in API mappings with a decorator as a target program element (\TotalDecoratorArgumentDeletionPercent).
On the other hand, \argTrans (purple) is common in API mappings with \fc as a source program element (\FunctionCallTotalArgumentTransformationPercent).
We observe that \DecoratorDecoratorArgumentNameChangePercent of \de to \de API mappings involve \argNC; however, none of the \de to \fc API mappings have \argNC.

\AsyncChange and \outTrans are only applicable to \fcs and \ats, but do not occur frequently (\TotalAsyncTransformationPercent and \TotalOutputTransformationPercent respectively of all API mappings).
We observe \ParamAddToDecorate even less frequently (\TotalParameterAdditionToDecoratedFunctionPercent of all API mappings).
Recall that we do not label properties for \imp code changes. Also, there are no properties when an API is just added or removed. Therefore there are no charts in the \imp and none rows and columns.

\begin{findingenv}{\ref{rq:code-changes}}{finding:rq1}
The majority of library pairs (\FunctionCallFunctionCallLPPercent) have API mappings between \fcs.
However, we also find that \HasNonFunctionLPPercent of library pairs have API mappings that involve non-function program elements.
Regardless of the type of program elements in an API mapping, \TotalElementNameChangePercent of API mappings involve program elements with \textit{different} names, and \FunctionCallFunctionCallArgumentTransformationPercent of \fc to \fc API mappings involve \argTrans.
\end{findingenv}

\subsubsection{\ref{rq:code-changes} discussion}
To the best of our knowledge, most of the current API mapping or client code transformation techniques consider \fc{s} as the main program element to find mappings for \cite{teyton2013automatic, alrubaye2018automating, alrubaye2019use, alrubaye2020learning, ni2021soar}.
If we assume these techniques work for Python, then the good news is that this implies that they may be able to support some API mappings in \FunctionCallFunctionCallLPPercent of the library pairs in our data that have \fc to \fc API mappings.
However, most of these techniques do \textit{not} consider other program element types (e.g., attributes or decorators), and worse, do not try to deduce mappings between different types of program elements \cite{teyton2013automatic, alrubaye2018automating, alrubaye2019use, alrubaye2020learning, ni2021soar}.
Our findings show that \HasNonFunctionLPPercent of library pairs require replacements that involve non-function program elements.

Most of the previous library migration work focuses on Java \cite{alrubaye2018automating,alrubaye2019migrationminer,teyton2013automatic}, which does not support the async/await style programming.
Supporting the \asyncChange we found in Python migrations would require adaptations to these techniques.

\subsection{\ref{rq:combination} \RQCombination}
\subsubsection{Motivation}  In \ref{rq:code-changes}, we analyze the program elements and properties involved in the API mappings we find.
However, a migration may require various code changes related to different API mappings.
To fully understand the type of tool support needed for observed migrations, we must also examine the combinations of code changes in a complete migration.
For example, our findings in \ref{rq:code-changes} reveal that \FunctionCallFunctionCallLPPercent of libraries have API mappings that include \fc as both source and target.
However, the associated migrations may also include code changes related to \ats.
Thus, in this RQ, we look at \textit{migrations} in terms of combinations of the code changes that occur.

\subsubsection{Methods}
For each migration, we initially identify the unique combinations of code changes in terms of their program elements and properties.
For instance, the migration depicted in \autoref{fig:1d8923a} has four code changes:
code changes \ccLines{35}{36-37} and \ccLines{98}{100-101} have
\ccDesc{\fc}{\fc}{\encShort, \asyncChangeShort, \argAddShort}
and code changes \ccLines{36}{38} and \ccLines{99}{102} have 
\ccDesc{\at}{\fc}{\encShort, \asyncChangeShort}
Therefore, the distinct combination of code changes in this example is {(1) 
\ccDesc{\fc}{\fc}{\encShort, \asyncChangeShort, \argAddShort}
and (2) 
\ccDesc{\at}{\fc}{\encShort, \asyncChangeShort}.
Next, for each unique combination of code changes, we determine the number of migrations that exhibit this combination.
Through this analysis, we identify a total of \UniqueCCCombo unique combinations (the full list is included in our artifacts).

\autoref{tab:mig-combinations} displays the 12 most frequent combinations with at least 3 migrations, which we discuss in the paper.
We note how row 11 shows a combination that has two different types of code changes, due to the properties.
Since an import statement change is usually required for all migrations, we ignored import code changes when determining the combinations.
This is why the \autoref{tab:mig-combinations} shows a total of 256 migrations instead of \VTwoMigCount as indicated in \autoref{tab:dataset}.
For the remaining 79 migrations, the only changes that occurred in the migration were changes to the import statements.

\subsubsection{Findings}

\begin{table*}[t]
    \centering
    \caption{Code change combinations in the migrations.}
    \label{tab:mig-combinations}
    \resizebox{1\textwidth}{!}{
        \begin{tabular}{@{}rlrrcrlrr@{}}
            \cmidrule{1-4}\cmidrule{6-9}
            \textbf{rank} & \textbf{combination}                                               & \multicolumn{1}{l}{\textbf{\# migs}} & \multicolumn{1}{l}{\textbf{\% migs}} &  &  \textbf{rank}        & \textbf{combination}                                                                                                                                  & \multicolumn{1}{l}{\textbf{\# migs}} & \multicolumn{1}{l}{\textbf{\% migs}} \\ \cmidrule{1-4}\cmidrule{6-9}
            1             & function call $\rightarrow$~ function call | elemNC                 & 21                                   & 8.2\%                                &  &  8                    & function call $\rightarrow$~ function call | argAdd, elemNC                                                                                            & 4                                    & 1.6\%                                \\ \arrayrulecolor{black!30}\cmidrule{1-4}\cmidrule{6-9}
            2             & function call $\rightarrow$~ function call | no properties          & 18                                   & 7.0\%                                &  &  9                    & function call $\rightarrow$~ function call | argTrans                                                                                                  & 4                                    & 1.6\%                                \\ \arrayrulecolor{black!30}\cmidrule{1-4}\cmidrule{6-9}
            3             & function call $\rightarrow$~ function call | argTrans, elemNC       & 11                                   & 4.3\%                                &  &  10                   & decorator $\rightarrow$~ decorator | argTrans                                                                                                          & 3                                    & 1.2\%                                \\ \arrayrulecolor{black!30}\cmidrule{1-4}\cmidrule{6-9}
            4             & function call $\rightarrow$~ function call | argAdd, argDel, elemNC & 8                                    & 3.1\%                                &  &  11                   & \begin{tabular}[c]{@{}l@{}}function call $\rightarrow$~ function call | elemNC\\ function call $\rightarrow$~function call | no properties\end{tabular}& 3                                    & 1.2\%                                \\ \arrayrulecolor{black!30}\cmidrule{1-4}\cmidrule{6-9}
            5             & function call $\rightarrow$~ function call | elemNC, outTrans       & 4                                    & 1.6\%                                &  &  12                   & function call $\rightarrow$~ function call | argDel                                                                                                    & 3                                    & 1.2\%                                \\ \arrayrulecolor{black!30}\cmidrule{1-4}\cmidrule{6-9}
            6             & function call $\rightarrow$~ function call | argDel, elemNC         & 4                                    & 1.6\%                                &  &  \multicolumn{1}{l}{} & remaining 155 combinations                                                                                            & 169                           & 66\%                                                                     \\ \arrayrulecolor{black!30}\cmidrule{1-4}\arrayrulecolor{black}\cmidrule{6-9}
            7             & function call $\rightarrow$~ function call | argAdd                 & 4                                    & 1.6\%                                &  &  \multicolumn{1}{l}{} & \textbf{total (167 combinations)}                                                                                                                     & \textbf{256}                         & \textbf{100\%}                       \\ \arrayrulecolor{black}\cmidrule{1-4}\cmidrule{6-9}
        \end{tabular}
    }
\end{table*}

From \autoref{tab:mig-combinations}, the most frequent combination is \fc replacements with \enc (\ComboFunctionCallWithElemNCPercent) followed by those with no properties (7\%).
Within the \fc replacements, we find several repeated property combinations, such as row 3 ``\argTransShort, \encShort'' (4.3\%), row 4 ``\argAddShort, \argDelShort, \encShort'' (3.1\%), and 5 others.
Overall, we observe 62 different combinations of properties associated with migrations that involve \fc replacement, which suggests that performing the \transform requires intricate client code changes.


While \HasFcComboCount of the \UniqueCCCombo unique combinations (\HasFcComboPercent) involve \fcs in some way, \ComboMixFcNonFcMigPercent of the migrations combine functions with other program elements, whether in the same code change or in a different code change. 
In \AttributeWithFCInSameCCPercent of all migrations, we observe \ats combined with \fc{s} within the same code change, while in \AttributeWithFCInDiffCCPercent migrations, \fc and \at appear in the same migration but separate code changes.
Similarly, \des pair with \fc in the same code change in \DecoratorWithFCInSameCCPercent of migrations, whereas \DecoratorWithFCInDiffCCPercent of migrations have \des in a different code change.
While \ex, \ty, and \ca appear with \fcs in \ExceptionWithFCInSameCCPercent, \TypeWithFCInSameCCPercent, and \FuncRefWithFCInSameCCPercent migrations, respectively, they do not appear in the same code change.
Overall, \ComboJustFcMigPercent migrations involve only \fcs, \ComboJustNonFCMigPercent migrations involve no \fcs, and the remaining \ComboMixFcNonFcMigPercent involve \fcs along with other program elements.

\begin{findingenv}{\ref{rq:combination}}{finding:rq2}
    44\% migrations are non-trivial, involving code changes with more than one type of program element and a variety of properties.
    Overall, only 16.4\% migrations involve just \fcs without any argument or output modifications.    
\end{findingenv}

\subsubsection{\ref{rq:combination} discussion}
Most existing library migration techniques focus on \fc{s} as the APIs to migrate \cite{teyton2013automatic, alrubaye2018automating, alrubaye2019use,ni2021soar}.
Conceptually, these would not be able to fully support \ComboHasNonFCMigPercent of the \migbenchTwo migrations with non-function program elements.
Numerous techniques identify only function API mappings, neglecting modifications in arguments or outputs \cite{teyton2013automatic, alrubaye2018automating, alrubaye2019use}.
Only 16.4\% of observed migrations had simple function call replacements without additional properties.

\subsection{\ref{rq:effort}. \RQEffort}
\subsubsection{Motivation} While RQ1 and RQ2 offer insights into the typical code changes required for library migration, they do not provide information on the associated development effort. 
In this RQ, we analyze our data to estimate the development effort needed for library migration, whether undertaken by a developer or an automated tool.

\begin{figure}
    \centering
    \includegraphics[width=\diffWidth\columnwidth]{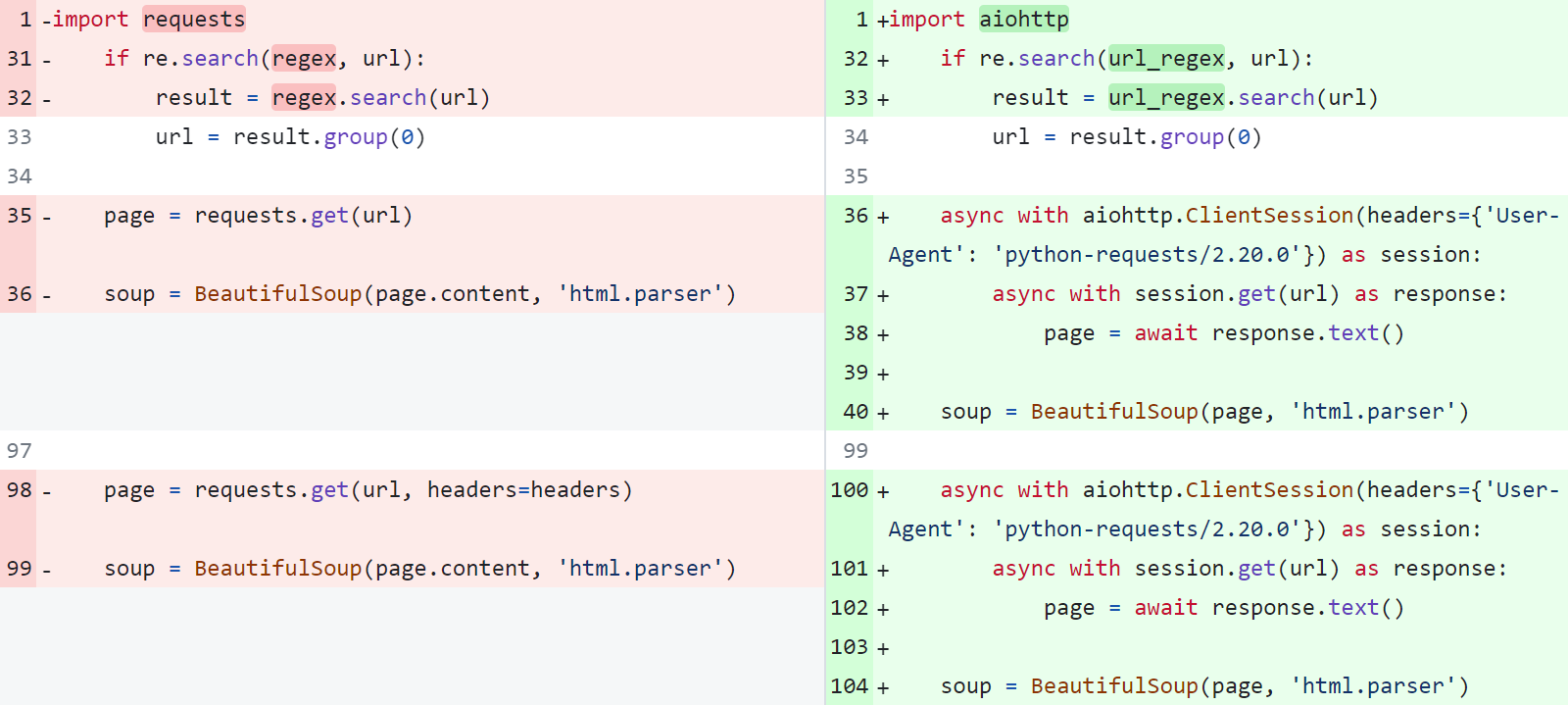}

    {\footnotesize $MigLOC=10, NumAPIs=10, NumChanges=4, UniqueAPIs=5, UniqueMappings=2$}
    \caption{
        \migDesc{requests}{aiohttp}{raptor123471/dingolingo}{1d8923a}        
    }
    \label{fig:1d8923a}
    \Description{}
\end{figure}

\subsubsection{Methods}
\label{sec:rq-migsize-method}
Although various metrics exist in software engineering \cite{metricsbook}, to the best of our knowledge, none directly measure the development effort required for a migration.
Drawing inspiration from metrics used to characterize \textit{code changes} in general~\cite{benestad2009understanding}, we define five simple metrics to estimate migration development effort.
Specifically, our collected migration data allows us to measure the development effort from two perspectives: the migration size and the minimum amount of APIs developers need to learn for a migration.

\begin{enumerate}[label=\textbf{M\arabic*},leftmargin=*]

\item \textit{MigLOC} is the sum of added and removed migration-related lines in non-import code changes in a migration.
It is an estimation of the size of a migration.
The diff in \autoref{fig:1d8923a}, has 7 removed lines (\remLine{1, 31-32, 35-36, 98-99}) and 13 added lines (\addLine{1, 32-33, 36-40, 100-104}).
Lines \remLine{31-32}, \addLine{32-33}, \addLine{39-40} and \addLine{103-104} are not migration related and lines \remLine{1} and \addLine{1} are import changes.
Therefore, there are 4 removed and 6 added migration-related added lines, making MigLOC=10.

 \item \textit{NumAPIs} is the total number of source and target API usages that were removed and added, respectively, in a migration. 
This provides an indication of the development effort required to find source API usages and replace them with corresponding target APIs.
In \autoref{fig:1d8923a}, the hunk \ccLines{35-36}{36-40} uses two source APIs \texttt{get()} and \texttt{content} and three target APIs \texttt{ClientSession()}, \texttt{get()}, and \texttt{text()}, totalling to 5 API usages.
The same API uses are repeated in hunk \ccLines{98-99}{100-104}, making NumAPIs for this diff = 5+5 =10.

 \item \textit{NumChanges} is the total number of code changes in a migration.
To replace source API usages, developers also need to understand the context (client code) in which the API has been used, which takes effort. Code changes in a migration can be intermittent, which means developers need to understand different contexts. NumChanges provides an indication of the effort required to understand the context. 
In the example in \autoref{fig:1d8923a}, there are a total of 4 non-import \ccs ( \ccLines{35}{36-37}, \ccLines{36}{40}, \ccLines{98}{100-101} and \ccLines{99}{102}).
, making NumChanges=4 for this diff.

\item\textit{ UniqueAPIs} is the number of \textit{unique} APIs from the source and the target library used in a migration.
When applying a migration, the developer has to understand the source and target APIs involved in the migration.
Accordingly, \textit{UniqueAPIs} is a proxy for the learning effort required for the migration. 
In \autoref{fig:1d8923a}, hunk \ccLines{35-36}{36-40} uses 2 source and 3 target APIs, which are repeated in hunk \ccLines{98-99}{100-104}, making UniqueAPIs=5 for this migration.

\item
\textit{UniqueMappings}
Other than the source and target APIs themselves, developers also have to learn which source API(s) should be replaced by which target API(s), \ie, API mapping.
We use \textit{UniqueMappings} to count the number of unique API mappings in a migration.
The migration in \autoref{fig:1d8923a} has a total of 4 non-import code changes: 
\ccLines{35}{36-37}, \ccLines{36}{38}, \ccLines{98}{100-101} and 
\ccLines{99}{102}.
However, both \ccLines{35}{36-37} and \ccLines{98}{100-101} replace the same function calls.
Similarly, code changes \ccLines{36}{38} and \ccLines{99}{102} are also identical.
Therefore, UniqueMappings=2 in this migration.
\end{enumerate}

\subsubsection{\ref{rq:effort} Findings}
\begin{figure*}[t]
    \centering
    \includegraphics[width=\textwidth]{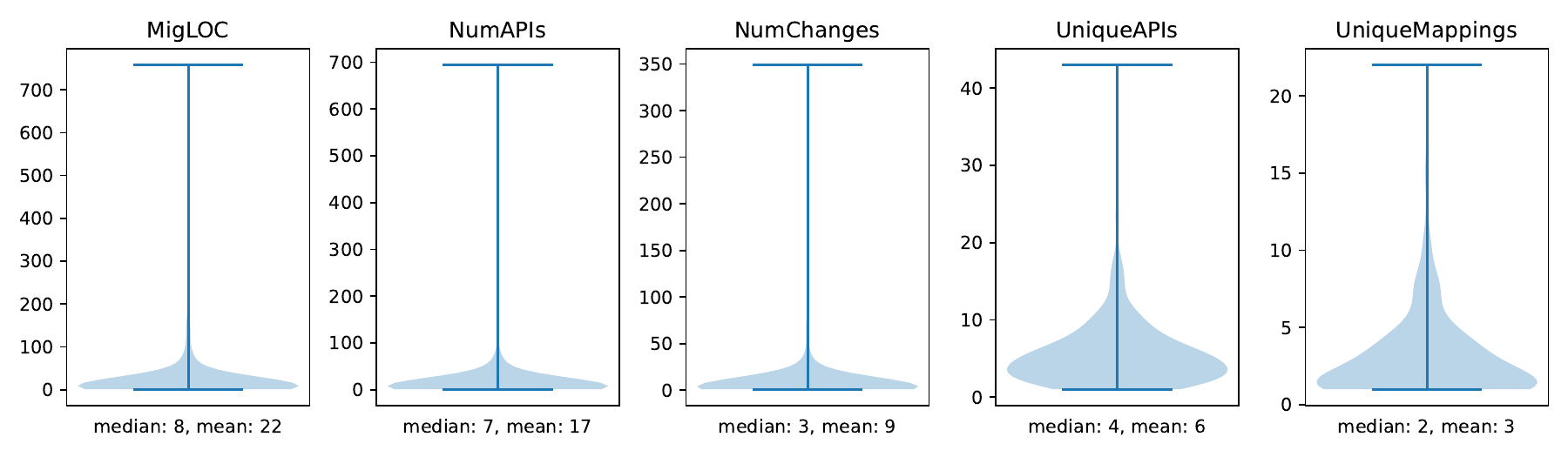}    
    \caption{Distribution of estimations of development effort.}
    \label{fig:effort}
    \Description{}
\end{figure*}

\autoref{fig:effort} shows the distribution of the five metrics.
We use the medians of these distributions to describe a typical migration.
On average, a migration involves \MigLOCMedian changed lines of code, \NumAPIsMedian API instances, and \NumChangesMedian code changes, representing the development effort.
From a learning standpoint, a developer typically needs to learn \UniqueAPIsMedian APIs and \UniqueMappingsMedian API mappings per migration.
We find that for all metrics, the mean values are much higher than the corresponding median value.
This means that the distributions are positively skewed, \ie, there are some migrations requiring very high development effort.
Recall that our dataset includes only single-commit migrations, suggesting the displayed values already represent a lower bound.

Some migrations may demand high development effort but not necessarily high learning effort. For example, the migration from library \lib{umsgpack} to \lib{msgpack} in commit \commitLinkPlain{logicaldash/lise}{028d0b34} involves substantial code changes (MigLOC=116, NumAPIs=72, NumChanges=36) but only necessitates learning 6 unique APIs and 3 mappings, indicating many repeated changes for the same mapping.
Conversely, other migrations, like from library \lib{twitter} to \lib{tweepy} in commit \commitLinkPlain{cloudbotirc/cloudbot}{f8243223}, necessitate both high development and learning effort due to larger differences between the libraries, requiring learning 35 unique APIs and 17 mappings.

\begin{findingenv}{\ref{rq:effort}}{finding:rq3}
   A typical Python library migration requires changing 8 LOC and 7 API instances, as well as understanding 4 APIs and 2 API mappings. However, some library migrations require higher number of lines of code and API changes as well as API understanding and mapping, reaching up to changing 758 lines of code, understanding 43 APIs, and mapping 22 APIs.    
\end{findingenv}

\subsubsection{\ref{rq:effort} Discussion}
The good news for both developers and tools is that the average migration does not require significant development and learning effort. However, some migrations do require considerable effort, highlighting the need for tool support to alleviate the manual burden of migrating libraries. Existing API mapping tools and techniques~\cite{miningAnalogicalAPIs, ni2021soar} can reduce learning effort.
However, as pointed above, comprehensive \transform tools are still missing.
That said, not all migrations with high development effort necessarily require high learning effort, it could be that there are many repeated code transformations in a migration.
However, such changes are still not simple ``find and replace'' or a well-defined refactoring that a developer can apply on a whole file.
One idea could be to develop tools that observe the developer as they make one of these changes and infer the change pattern that could be applied to the rest of the file (e.g., something similar to learning from examples~\cite{ketkar2022inferring}).
In fact, it would be interesting to see if Large Language Models already integrated in the IDE, e.g. Copilot, might be effective at such a task.

\section{Implications}
\label{sec:implication}
The contributions of this paper as well as the findings from our empirical study have several implications for library migration researchers and tool builders.

\subsection{Implications for API mapping research}
\migbenchTwo, unlike the original \migbench, includes APIs from each of the \VTwoCCCount manually verified code changes in the dataset.
This serves as a ground truth, aiding researchers in evaluating and training their API mapping techniques.
as well as train their techniques using this data. While most of the existing API mapping techniques focus on one-to-one API mapping, \migbenchTwo includes code changes that involve  higher cardinality API replacements. This indicates a need for higher cardinality API mapping techniques, which \migbenchTwo can facilitate.  

\subsection{Implications for client code transformation research}
As discussed in RQ2, we find that the \transform{s} required in real migrations go beyond only identifying API mappings, and also beyond focusing on mappings between the same types of program elements.
From RQ1, we also find that some migrations require taking synchronizations into account, especially if one library supports asynchronous programming while the other does not.
Migration tools that can automatically infer the need to add \texttt{async}/\texttt{await} keywords are needed.
Overall, our characterization of \codechange{s} and analysis of their frequency in typical migrations allow tool developers to make informed decisions about the type of \transform{} they should support.
Additionally, the labeled and validated \ccs{} we provide in \migbenchTwo also facilitate the evaluation of any developed migration techniques.

\subsection{Implications for systematic labeling and comparisons}
Researchers can use \taxonomy to label the capabilities of their techniques and fairly compare library migration techniques and identify their limitations.
For example, as a conceptual labeling, SOAR \cite{ni2021soar} supports \oo and \om{} \fc replacements along with function name change and argument addition, deletion, and transformation.
However, SOAR does not support \des, \ats, \ty{s} and \mo or \mm{} \fc replacements present in \taxonomy.
Moreover, SOAR does not support \asyncChange, which means it cannot migrate between synchronous and asynchronous libraries.  
Together, \taxonomy and \migbenchTwo allow both conceptual and empirical comparisons of library migration techniques.
This allows the identification of limitations of existing techniques and then developing new techniques to address these limitations.

\section{Threats to Validity}
\label{sec:threats}
\subsection{Internal validity}
To enhance the accuracy and correctness of \salm and reduce manual effort in identifying \ccs, we performed some automated steps. We used OpenAI's GPT-4 API to identify non-analogous libraries and to extract libraries' import names. The results produced by GPT-4 may not be completely reliable.
We manually verified 25 (5\% of all) of GPT-4’s results and 21 of 25 (84\%) were correct.
GPT-4 marked the remaining four as non-analogous, but we found them to be analogous.
We acknowledge that if GPT-4 incorrectly indicates that two libraries are non-analogous, we will miss the related migrations completely.
That said, our chosen methodology matches our goal of ensuring that migrations that end up in our benchmark are actually migrations.

We also ignored the libraries that are not available in PyPI and their corresponding migrations, because we focus on only third-party libraries. In all these steps, we may have filtered out true migrations between analogous third-party libraries and thus missed their \ccs.
To minimize this, we validated import names and ignored migrations without \ccs during manual review.
Since our manually verified migrations from \salm cover \SALMLabeledDomainPercent of the library domains and \SALMLabeledLPPercent of the library pairs from the original dataset, it is unlikely that any overlooked migrations would significantly differ from our observations.
This assumption is supported by reaching data saturation after labeling 52 migrations from \salm.

In identifying \ccs, we considered only visible modifications detected by git's diff, thereby missing instances where an API is replaced with an identical one (e.g., the same qualified name and signature for function calls).
While this could have potentially increased the percentage of program elements with no element name change, we deem such cases as less relevant for library migration research, given that identical APIs require no migration action.

In a migration, a source library could be replaced by many target libraries and vice versa. In this study, we assume that one library is replaced by one one target library. Although this does not affect our taxonomy, considering multiple source or target libraries may have increased the estimated development effort. Thus, our results are primarily applicable to one-to-one library migrations.

\subsection{Construct validity}
We manually review migrations to identify and label \ccs, relying on the authors’ library knowledge, which could potentially lead to mislabeling.
To mitigate this, we reviewed each library’s documentation and examples, ensuring sufficient understanding.
We erred on the side of caution and skipped unclear migrations (e.g., commits with too many tangled changes and refactorings), clearly marking them as such.
We also refined our results via several iterations of
discussions among the authors and had two authors independently review each migration until achieving substantial inter-rater agreement and data saturation.

The decision to categorize a set of APIs involved in a migration as multiple \oo \ccs or one higher-cardinality \cc can often be subjective.
We use the API usages, names, arguments, and documentation to determine if \oo relations can be established.
In cases of ambiguity, we consistently categorize these as higher cardinality changes.

Our work focuses on migration-related changes rather than stylistic ones, so we do not consider changes in the qualified name as \ccs during labeling; these can be inferred from the function name and import statement. Style-related changes were ignored in the identification and labeling process. All labeled migration data is available for external review and verification.

The metrics we use in RQ3 reflect various characteristic of a migration as \textit{proxies} for development effort, and as inspired by existing software engineering and software evolution principles/metrics.
We acknowledge these characteristics or metrics haven’t been validated to directly correlate with perceived migration development effort.
Additionally, we can only approximate the effort related to observable code changes, without information on testing or additional discussions between team members.
Our aim is to provide some \textit{perspective} of migration effort, with validating and improving these metrics being an interesting avenue for future work.

\subsection{External validity}
The datasets we use only consider single-commit migrations, but previous studies have shown that library migration can span multiple commits \cite{teyton2013automatic}.
For this first effort on systematically characterizing Python library migrations, we limited ourselves to existing datasets.
In the future, we hope to build multi-commit Python migration datasets and investigate whether multi-commit migrations are more complex than single-commit migrations.


\taxonomy is based on code changes between \VTwoLPCount \libpairs containing \VTwoLibCount unique libraries across \VTwoDomainCount domains.
While we aimed for data saturation using two datasets and followed a stratified sampling approach to analyze a statistically representative sample of migrations, we cannot guarantee that we observed all types of code changes.
We document our extraction and review process to allow others to extend \taxonomy and \migbenchTwo.

\section{Related work}
\label{sec:related-work}

\subsection{Library migration datasets and benchmarks}
Teyton et al. \cite{teyton2012mining,teyton2014study} and He et al. \cite{he2021multi,he2021large} provide benchmarks for Java library migration. Teyton et al.'s initial work \cite{teyton2012mining} developed a semi-automatic approach to detect 80 analogous Java library pairs by mining Maven repository histories and manually verifying the results. They later refined their methodology \cite{teyton2014study} to capture multi-commit migrations, ultimately identifying 329 \libpairs across 32 domains. He et al. \cite{he2021multi} took a different approach by using both Teyton et al.'s mining technique and their own recommendation system to validate a dataset comprising 1,434 analogous \libpairs. Later, they expanded this to include 3,163 manually verified migration commits \cite{he2021large}. Both efforts share a common approach in terms of mining client repository history, filtering candidates based on specific metrics (e.g., selecting repositories with more than 10 stars, or analyzing the project dependency files), and manual verification.
These collectively provide a robust methodological foundation for mining library migrations.
These methods were also followed by the Python library migration datasets we use in this paper, SALM~\cite{salm} and \migbench~\cite{migbench}, and which we discussed in detail in Section~\ref{subsec:datasets}.
Our work relies on SALM and \migbench, but further expands these datasets by identifying, labeling, and categorizing \codechanges. Notably, aside from line numbers in \migbench, neither of the original datasets contains labeled code changes.

\subsection{API mapping}
API mapping techniques for library migration can be grouped into history-based and non-history-based approaches. \textit{History-based methods}, represented by Teyton et al. \cite{teyton2013automatic} and Alrubaye et al. \cite{alrubaye2018automating}, analyze migration commits to discern and filter function mappings, addressing issues like higher cardinality mappings \cite{alrubaye2019use} and incorporating machine learning from API documentation \cite{alrubaye2020learning}.
In contrast, \textit{non-history-based techniques}, like the work by Chen et al. \cite{miningAnalogicalAPIs} and SOAR \cite{ni2021soar}, employ unsupervised deep learning models or textual similarity to identify API mappings. Non-history based techniques are especially useful when there are insufficient historical migration examples.
Despite their merits, both techniques have limitations: history-based approaches struggle with new libraries \cite{ni2021soar}, SOAR, requiring ample documentation, has been tested on only two domain-specific libraries (R and Python)~\cite{aghajani2020software, ni2021soar}.

\subsection{Client code transformation}
The ultimate aim of library migration research is to develop tools that can automatically transform client code that uses one library to use a different library. Balaban \etal \cite{balaban2005refactoring} developed a technique to migrate the uses of legacy Java classes. SOAR above \cite{ni2021soar} used program synthesis techniques for client code transformation based on identified API mappings, though it is been evaluated on just one Python library pair. SOAR is notably the only known technique for Python.
Additionally, there is a substantial body of work on version migration and updating client code due to breaking changes, such as \cite{xing2007api,nielsen2021semantic,AlineAPIDiff18}, addressing a related issue.
While our study does not propose automated techniques, our labeling of \codechanges helps identify potential API mappings between analogous Python libraries.
This lays foundational knowledge for future API mapping and code transformation methods, offering insights into the types of code changes to be addressed and supplying labeled benchmark data for evaluation.

\subsection{Library migration taxonomies}
To our knowledge, no standard taxonomy exists for types of migration-related code changes.
While some techniques presented above \cite{teyton2013automatic, alrubaye2018automating,alrubaye2019use,alrubaye2020learning,ni2021soar} address various code changes, there is no systematic analysis of possible change types, making their selection unclear.
For instance, Alrubaye et al. categorize API mappings into one-to-one, one-to-many, and many-to-many but only discuss functions. SOAR \cite{ni2021soar} considers the need for argument addition, deletion, and transformation but but is limited by the target library/domain.
Our systematic approach to understanding Python \ccs revealed previously unconsidered changes, such as replacements across different types of program elements).

\subsection{Measuring migration complexity and effort}
Defining reliable metrics in software engineering is a long-time challenge, with numerous proposed metrics to measure code complexity~\cite{metricsbook,abran2010software,mccabe1976complexity,halstead1977elements, hariprasad2017software, curtis1979measuring}, or qualitative levels of complexity~\cite{boehm2000software}.
There has also been research been devoted to estimating development effort~\cite{albrecht1983software,gautam2018state,jorgensen2014we,hill2011practical,sharma2012estimation,mahmood2020systematic}. 
While some methods, like \cite{albrecht1983software}, use code-based metrics, most analyze software process, relying on requirements specifications, use cases, work breakdown structure, and other artifacts, often combined with expert knowledge.
Our characterization of migration development effort through size related metrics and proxies for learning effort were inspired by the above literature as well as code change measurements~\cite{benestad2009understanding}.

\section{Conclusion}
\label{sec:conclusion}
We conducted an empirical study to gain a comprehensive understanding of Python library migrations.
To enable this empirical study, we contributed \migbenchTwo, a manually curated library migration dataset,
and \taxonomy, a taxonomy for categorizing \codechanges.
\migbenchTwo comprises a collection of \VTwoMigCount migrations sourced from two state-of-the-art migration datasets, 
which we have significantly enhanced by incorporating a \VTwoCCCount manually verified and labeled \codechange instances.

From our empirical study, we find that \HasNonFunctionLPPercent of library pairs have API mappings that involve non-function program elements.
We find that 16.4\% of the migrations are trivial, involving only \fcs replacements without any argument or output modifications.
However, a larger proportion (44\%) are non-trivial, involving code changes with more than one type of program element and a variety of properties.
As an approximation for the development effort involved, we find that, on average, a developer needs to learn about \UniqueAPIsMedian APIs and \UniqueMappingsMedian API mappings to perform a migration, and change \MigLOCMedian lines of code.
However, we also found cases of migrations that involve up to \UniqueAPIsMax unique APIs, \UniqueMappingsMax API mappings, and \MigLOCMax lines of code, making them harder to manually implement.
Overall, our contributions provide the necessary knowledge and foundations for developing and evaluating automated Python library migration techniques.

\section{Data Availability}
\label{sec:data-availability}
Replication package available at \artifactURL.

\section*{acknowledgement}
We acknowledge the support of the Natural Sciences and Engineering Research Council of Canada (NSERC) through their CREATE, Discovery, and Canada Research Chairs programs.

\bibliographystyle{ACM-Reference-Format}
\bibliography{references}   


\begin{thebibliography}{59}


\ifx \showCODEN    \undefined \def \showCODEN     #1{\unskip}     \fi
\ifx \showDOI      \undefined \def \showDOI       #1{#1}\fi
\ifx \showISBNx    \undefined \def \showISBNx     #1{\unskip}     \fi
\ifx \showISBNxiii \undefined \def \showISBNxiii  #1{\unskip}     \fi
\ifx \showISSN     \undefined \def \showISSN      #1{\unskip}     \fi
\ifx \showLCCN     \undefined \def \showLCCN      #1{\unskip}     \fi
\ifx \shownote     \undefined \def \shownote      #1{#1}          \fi
\ifx \showarticletitle \undefined \def \showarticletitle #1{#1}   \fi
\ifx \showURL      \undefined \def \showURL       {\relax}        \fi
\providecommand\bibfield[2]{#2}
\providecommand\bibinfo[2]{#2}
\providecommand\natexlab[1]{#1}
\providecommand\showeprint[2][]{arXiv:#2}

\bibitem[Abdalkareem et~al\mbox{.}(2017)]%
        {trivialNPMPackage}
\bibfield{author}{\bibinfo{person}{Rabe Abdalkareem}, \bibinfo{person}{Olivier
  Nourry}, \bibinfo{person}{Sultan Wehaibi}, \bibinfo{person}{Suhaib Mujahid},
  {and} \bibinfo{person}{Emad Shihab}.} \bibinfo{year}{2017}\natexlab{}.
\newblock \showarticletitle{Why do developers use trivial packages? an
  empirical case study on npm}. In \bibinfo{booktitle}{\emph{Proceedings of the
  2017 11th joint meeting on foundations of software engineering}}.
  \bibinfo{pages}{385--395}.
\newblock


\bibitem[Abran(2010)]%
        {abran2010software}
\bibfield{author}{\bibinfo{person}{Alain Abran}.}
  \bibinfo{year}{2010}\natexlab{}.
\newblock \bibinfo{booktitle}{\emph{Software metrics and software metrology}}.
\newblock \bibinfo{publisher}{John Wiley \& Sons}.
\newblock


\bibitem[Aghajani et~al\mbox{.}(2020)]%
        {aghajani2020software}
\bibfield{author}{\bibinfo{person}{Emad Aghajani}, \bibinfo{person}{Csaba
  Nagy}, \bibinfo{person}{Mario Linares-V{\'a}squez}, \bibinfo{person}{Laura
  Moreno}, \bibinfo{person}{Gabriele Bavota}, \bibinfo{person}{Michele Lanza},
  {and} \bibinfo{person}{David~C Shepherd}.} \bibinfo{year}{2020}\natexlab{}.
\newblock \showarticletitle{Software documentation: the practitioners'
  perspective}. In \bibinfo{booktitle}{\emph{2020 IEEE/ACM 42nd International
  Conference on Software Engineering (ICSE)}}. IEEE, \bibinfo{pages}{590--601}.
\newblock


\bibitem[Albrecht and Gaffney(1983)]%
        {albrecht1983software}
\bibfield{author}{\bibinfo{person}{Allan~J. Albrecht} {and}
  \bibinfo{person}{John~E Gaffney}.} \bibinfo{year}{1983}\natexlab{}.
\newblock \showarticletitle{Software function, source lines of code, and
  development effort prediction: a software science validation}.
\newblock \bibinfo{journal}{\emph{IEEE transactions on software engineering}}
  \bibinfo{number}{6} (\bibinfo{year}{1983}), \bibinfo{pages}{639--648}.
\newblock


\bibitem[Alrubaye and Mkaouer(2018)]%
        {alrubaye2018automating}
\bibfield{author}{\bibinfo{person}{Hussein Alrubaye} {and}
  \bibinfo{person}{Mohamed~Wiem Mkaouer}.} \bibinfo{year}{2018}\natexlab{}.
\newblock \showarticletitle{Automating the detection of third-party Java
  library migration at the function level.}. In
  \bibinfo{booktitle}{\emph{CASCON}}. \bibinfo{pages}{60--71}.
\newblock


\bibitem[Alrubaye et~al\mbox{.}(2020)]%
        {alrubaye2020learning}
\bibfield{author}{\bibinfo{person}{Hussein Alrubaye},
  \bibinfo{person}{Mohamed~Wiem Mkaouer}, \bibinfo{person}{Igor Khokhlov},
  \bibinfo{person}{Leon Reznik}, \bibinfo{person}{Ali Ouni}, {and}
  \bibinfo{person}{Jason Mcgoff}.} \bibinfo{year}{2020}\natexlab{}.
\newblock \showarticletitle{Learning to recommend third-party library migration
  opportunities at the API level}.
\newblock \bibinfo{journal}{\emph{Applied Soft Computing}}
  \bibinfo{volume}{90} (\bibinfo{year}{2020}), \bibinfo{pages}{106140}.
\newblock


\bibitem[Alrubaye et~al\mbox{.}(2019a)]%
        {alrubaye2019migrationminer}
\bibfield{author}{\bibinfo{person}{Hussein Alrubaye},
  \bibinfo{person}{Mohamed~Wiem Mkaouer}, {and} \bibinfo{person}{Ali Ouni}.}
  \bibinfo{year}{2019}\natexlab{a}.
\newblock \showarticletitle{Migrationminer: An automated detection tool of
  third-party java library migration at the method level}. In
  \bibinfo{booktitle}{\emph{2019 IEEE international conference on software
  maintenance and evolution (ICSME)}}. IEEE, \bibinfo{pages}{414--417}.
\newblock


\bibitem[Alrubaye et~al\mbox{.}(2019b)]%
        {alrubaye2019use}
\bibfield{author}{\bibinfo{person}{Hussein Alrubaye},
  \bibinfo{person}{Mohamed~Wiem Mkaouer}, {and} \bibinfo{person}{Ali Ouni}.}
  \bibinfo{year}{2019}\natexlab{b}.
\newblock \showarticletitle{On the use of information retrieval to automate the
  detection of third-party java library migration at the method level}. In
  \bibinfo{booktitle}{\emph{2019 IEEE/ACM 27th International Conference on
  Program Comprehension (ICPC)}}. IEEE, \bibinfo{pages}{347--357}.
\newblock


\bibitem[Balaban et~al\mbox{.}(2005)]%
        {balaban2005refactoring}
\bibfield{author}{\bibinfo{person}{Ittai Balaban}, \bibinfo{person}{Frank Tip},
  {and} \bibinfo{person}{Robert Fuhrer}.} \bibinfo{year}{2005}\natexlab{}.
\newblock \showarticletitle{Refactoring support for class library migration}.
\newblock \bibinfo{journal}{\emph{ACM SIGPLAN Notices}} \bibinfo{volume}{40},
  \bibinfo{number}{10} (\bibinfo{year}{2005}), \bibinfo{pages}{265--279}.
\newblock


\bibitem[Baltes and Ralph(2022)]%
        {baltes2022sampling}
\bibfield{author}{\bibinfo{person}{Sebastian Baltes} {and}
  \bibinfo{person}{Paul Ralph}.} \bibinfo{year}{2022}\natexlab{}.
\newblock \showarticletitle{Sampling in software engineering research: A
  critical review and guidelines}.
\newblock \bibinfo{journal}{\emph{Empirical Software Engineering}}
  \bibinfo{volume}{27}, \bibinfo{number}{4} (\bibinfo{year}{2022}),
  \bibinfo{pages}{94}.
\newblock


\bibitem[Bauer and Heinemann(2012)]%
        {understandingAPIUsage}
\bibfield{author}{\bibinfo{person}{Veronika Bauer} {and} \bibinfo{person}{Lars
  Heinemann}.} \bibinfo{year}{2012}\natexlab{}.
\newblock \showarticletitle{Understanding API usage to support informed
  decision making in software maintenance}. In \bibinfo{booktitle}{\emph{2012
  16th European Conference on Software Maintenance and Reengineering}}. IEEE,
  \bibinfo{pages}{435--440}.
\newblock


\bibitem[Benestad et~al\mbox{.}(2009)]%
        {benestad2009understanding}
\bibfield{author}{\bibinfo{person}{Hans~Christian Benestad},
  \bibinfo{person}{Bente Anda}, {and} \bibinfo{person}{Erik Arisholm}.}
  \bibinfo{year}{2009}\natexlab{}.
\newblock \showarticletitle{Understanding software maintenance and evolution by
  analyzing individual changes: a literature review}.
\newblock \bibinfo{journal}{\emph{Journal of Software Maintenance and
  Evolution: Research and Practice}} \bibinfo{volume}{21}, \bibinfo{number}{6}
  (\bibinfo{year}{2009}), \bibinfo{pages}{349--378}.
\newblock


\bibitem[Boehm and Reifer(2000)]%
        {boehm2000software}
\bibfield{author}{\bibinfo{person}{B Boehm} {and} \bibinfo{person}{D Reifer}.}
  \bibinfo{year}{2000}\natexlab{}.
\newblock \showarticletitle{Software Cost Estimation with COCOMO II. Prentice
  Hall}.
\newblock \bibinfo{journal}{\emph{Upper Saddle River, NJ}}
  (\bibinfo{year}{2000}).
\newblock


\bibitem[Brito et~al\mbox{.}(2018)]%
        {AlineAPIDiff18}
\bibfield{author}{\bibinfo{person}{Aline Brito}, \bibinfo{person}{Laerte
  Xavier}, \bibinfo{person}{Andre Hora}, {and} \bibinfo{person}{Marco~Tulio
  Valente}.} \bibinfo{year}{2018}\natexlab{}.
\newblock \showarticletitle{APIDiff: Detecting API breaking changes}. In
  \bibinfo{booktitle}{\emph{2018 IEEE 25th International Conference on Software
  Analysis, Evolution and Reengineering (SANER)}}. \bibinfo{pages}{507--511}.
\newblock
\urldef\tempurl%
\url{https://doi.org/10.1109/SANER.2018.8330249}
\showDOI{\tempurl}


\bibitem[Chen et~al\mbox{.}(2019)]%
        {miningAnalogicalAPIs}
\bibfield{author}{\bibinfo{person}{Chunyang Chen}, \bibinfo{person}{Zhenchang
  Xing}, \bibinfo{person}{Yang Liu}, {and} \bibinfo{person}{Kent Ong~Long
  Xiong}.} \bibinfo{year}{2019}\natexlab{}.
\newblock \showarticletitle{Mining likely analogical apis across third-party
  libraries via large-scale unsupervised api semantics embedding}.
\newblock \bibinfo{journal}{\emph{IEEE Transactions on Software Engineering}}
  \bibinfo{volume}{47}, \bibinfo{number}{3} (\bibinfo{year}{2019}),
  \bibinfo{pages}{432--447}.
\newblock


\bibitem[Chen et~al\mbox{.}(2021)]%
        {ChenAPIMapping2021}
\bibfield{author}{\bibinfo{person}{Chunyang Chen}, \bibinfo{person}{Zhenchang
  Xing}, \bibinfo{person}{Yang Liu}, {and} \bibinfo{person}{Kent Ong~Long
  Xiong}.} \bibinfo{year}{2021}\natexlab{}.
\newblock \showarticletitle{Mining Likely Analogical {APIs} Across Third-Party
  Libraries via Large-Scale Unsupervised API Semantics Embedding}.
\newblock \bibinfo{journal}{\emph{IEEE Transactions on Software Engineering}}
  \bibinfo{volume}{47}, \bibinfo{number}{3} (\bibinfo{year}{2021}),
  \bibinfo{pages}{432--447}.
\newblock
\urldef\tempurl%
\url{https://doi.org/10.1109/TSE.2019.2896123}
\showDOI{\tempurl}


\bibitem[Cohen(1960)]%
        {cohen1960coefficient}
\bibfield{author}{\bibinfo{person}{Jacob Cohen}.}
  \bibinfo{year}{1960}\natexlab{}.
\newblock \showarticletitle{A coefficient of agreement for nominal scales}.
\newblock \bibinfo{journal}{\emph{Educational and psychological measurement}}
  \bibinfo{volume}{20}, \bibinfo{number}{1} (\bibinfo{year}{1960}),
  \bibinfo{pages}{37--46}.
\newblock


\bibitem[Curtis et~al\mbox{.}(1979)]%
        {curtis1979measuring}
\bibfield{author}{\bibinfo{person}{Bill Curtis}, \bibinfo{person}{Sylvia~B.
  Sheppard}, \bibinfo{person}{Phil Milliman}, \bibinfo{person}{MA Borst}, {and}
  \bibinfo{person}{Tom Love}.} \bibinfo{year}{1979}\natexlab{}.
\newblock \showarticletitle{Measuring the psychological complexity of software
  maintenance tasks with the Halstead and McCabe metrics}.
\newblock \bibinfo{journal}{\emph{IEEE Transactions on software engineering}}
  \bibinfo{number}{2} (\bibinfo{year}{1979}), \bibinfo{pages}{96--104}.
\newblock


\bibitem[Dabic et~al\mbox{.}(2021)]%
        {dabic2021sampling}
\bibfield{author}{\bibinfo{person}{O. Dabic}, \bibinfo{person}{E. Aghajani},
  {and} \bibinfo{person}{G. Bavota}.} \bibinfo{year}{2021}\natexlab{}.
\newblock \showarticletitle{Sampling Projects in GitHub for MSR Studies}. In
  \bibinfo{booktitle}{\emph{2021 2021 IEEE/ACM 18th International Conference on
  Mining Software Repositories (MSR) (MSR)}}. \bibinfo{publisher}{IEEE Computer
  Society}, \bibinfo{address}{Los Alamitos, CA, USA},
  \bibinfo{pages}{560--564}.
\newblock
\urldef\tempurl%
\url{https://doi.org/10.1109/MSR52588.2021.00074}
\showDOI{\tempurl}


\bibitem[Derr et~al\mbox{.}(2017)]%
        {keepMeUpdated}
\bibfield{author}{\bibinfo{person}{Erik Derr}, \bibinfo{person}{Sven Bugiel},
  \bibinfo{person}{Sascha Fahl}, \bibinfo{person}{Yasemin Acar}, {and}
  \bibinfo{person}{Michael Backes}.} \bibinfo{year}{2017}\natexlab{}.
\newblock \showarticletitle{Keep me updated: An empirical study of third-party
  library updatability on android}. In \bibinfo{booktitle}{\emph{Proceedings of
  the 2017 ACM SIGSAC Conference on Computer and Communications Security}}.
  \bibinfo{pages}{2187--2200}.
\newblock


\bibitem[Fenton and Bieman(2014)]%
        {metricsbook}
\bibfield{author}{\bibinfo{person}{Norman Fenton} {and} \bibinfo{person}{James
  Bieman}.} \bibinfo{year}{2014}\natexlab{}.
\newblock \bibinfo{booktitle}{\emph{Software metrics: a rigorous and practical
  approach}}.
\newblock \bibinfo{publisher}{CRC press}.
\newblock


\bibitem[Foundation({[n.\,d.]})]%
        {pypi}
\bibfield{author}{\bibinfo{person}{Python~Software Foundation}.}
  \bibinfo{year}{[n.\,d.]}\natexlab{}.
\newblock \bibinfo{booktitle}{\emph{Python Package Index - PyPI}}.
\newblock
\urldef\tempurl%
\url{https://pypi.org}
\showURL{%
Retrieved March 31, 2022 from \tempurl}


\bibitem[Gautam and Singh(2018)]%
        {gautam2018state}
\bibfield{author}{\bibinfo{person}{Swarnima~Singh Gautam} {and}
  \bibinfo{person}{Vrijendra Singh}.} \bibinfo{year}{2018}\natexlab{}.
\newblock \showarticletitle{The state-of-the-art in software development effort
  estimation}.
\newblock \bibinfo{journal}{\emph{Journal of Software: Evolution and Process}}
  \bibinfo{volume}{30}, \bibinfo{number}{12} (\bibinfo{year}{2018}),
  \bibinfo{pages}{e1983}.
\newblock


\bibitem[Gousios and Spinellis(2012)]%
        {ghtorrent}
\bibfield{author}{\bibinfo{person}{Georgios Gousios} {and}
  \bibinfo{person}{Diomidis Spinellis}.} \bibinfo{year}{2012}\natexlab{}.
\newblock \showarticletitle{GHTorrent: Github's data from a firehose}. In
  \bibinfo{booktitle}{\emph{2012 9th IEEE Working Conference on Mining Software
  Repositories (MSR)}}. \bibinfo{pages}{12--21}.
\newblock
\urldef\tempurl%
\url{https://doi.org/10.1109/MSR.2012.6224294}
\showDOI{\tempurl}


\bibitem[Gu et~al\mbox{.}(2023)]%
        {salm}
\bibfield{author}{\bibinfo{person}{Haiqiao Gu}, \bibinfo{person}{Hao He}, {and}
  \bibinfo{person}{Minghui Zhou}.} \bibinfo{year}{2023}\natexlab{}.
\newblock \showarticletitle{Self-Admitted Library Migrations in Java,
  JavaScript, and Python Packaging Ecosystems: A Comparative Study}. In
  \bibinfo{booktitle}{\emph{2023 IEEE International Conference on Software
  Analysis, Evolution and Reengineering (SANER)}}. IEEE,
  \bibinfo{pages}{627--638}.
\newblock


\bibitem[Halstead(1977)]%
        {halstead1977elements}
\bibfield{author}{\bibinfo{person}{Maurice~H Halstead}.}
  \bibinfo{year}{1977}\natexlab{}.
\newblock \bibinfo{booktitle}{\emph{Elements of Software Science (Operating and
  programming systems series)}}.
\newblock \bibinfo{publisher}{Elsevier Science Inc.}
\newblock


\bibitem[Hariprasad et~al\mbox{.}(2017)]%
        {hariprasad2017software}
\bibfield{author}{\bibinfo{person}{T Hariprasad}, \bibinfo{person}{G
  Vidhyagaran}, \bibinfo{person}{K Seenu}, {and} \bibinfo{person}{Chandrasegar
  Thirumalai}.} \bibinfo{year}{2017}\natexlab{}.
\newblock \showarticletitle{Software complexity analysis using halstead
  metrics}. In \bibinfo{booktitle}{\emph{2017 international conference on
  trends in electronics and informatics (ICEI)}}. IEEE,
  \bibinfo{pages}{1109--1113}.
\newblock


\bibitem[He et~al\mbox{.}(2021a)]%
        {he2021large}
\bibfield{author}{\bibinfo{person}{Hao He}, \bibinfo{person}{Runzhi He},
  \bibinfo{person}{Haiqiao Gu}, {and} \bibinfo{person}{Minghui Zhou}.}
  \bibinfo{year}{2021}\natexlab{a}.
\newblock \showarticletitle{A large-scale empirical study on Java library
  migrations: prevalence, trends, and rationales}. In
  \bibinfo{booktitle}{\emph{Proceedings of the 29th ACM Joint Meeting on
  European Software Engineering Conference and Symposium on the Foundations of
  Software Engineering}}. \bibinfo{pages}{478--490}.
\newblock


\bibitem[He et~al\mbox{.}(2021b)]%
        {he2021multi}
\bibfield{author}{\bibinfo{person}{Hao He}, \bibinfo{person}{Yulin Xu},
  \bibinfo{person}{Yixiao Ma}, \bibinfo{person}{Yifei Xu},
  \bibinfo{person}{Guangtai Liang}, {and} \bibinfo{person}{Minghui Zhou}.}
  \bibinfo{year}{2021}\natexlab{b}.
\newblock \showarticletitle{A multi-metric ranking approach for library
  migration recommendations}. In \bibinfo{booktitle}{\emph{2021 IEEE
  International Conference on Software Analysis, Evolution and Reengineering
  (SANER)}}. IEEE, \bibinfo{pages}{72--83}.
\newblock


\bibitem[Hill(2011)]%
        {hill2011practical}
\bibfield{author}{\bibinfo{person}{Peter~R Hill}.}
  \bibinfo{year}{2011}\natexlab{}.
\newblock \bibinfo{booktitle}{\emph{Practical software project estimation: a
  toolkit for estimating software development effort \& duration}}.
\newblock \bibinfo{publisher}{McGraw-Hill Education}.
\newblock


\bibitem[Islam et~al\mbox{.}(2023)]%
        {migbench}
\bibfield{author}{\bibinfo{person}{Mohayeminul Islam},
  \bibinfo{person}{Ajay~Kumar Jha}, \bibinfo{person}{Sarah Nadi}, {and}
  \bibinfo{person}{Ildar Akhmetov}.} \bibinfo{year}{2023}\natexlab{}.
\newblock \showarticletitle{PyMigBench: A Benchmark for Python Library
  Migration}. In \bibinfo{booktitle}{\emph{2023 IEEE/ACM 20th International
  Conference on Mining Software Repositories (MSR)}}. IEEE,
  \bibinfo{pages}{511--515}.
\newblock
\urldef\tempurl%
\url{https://doi.org/10.1109/MSR59073.2023.00075}
\showDOI{\tempurl}


\bibitem[J{\o}rgensen(2014)]%
        {jorgensen2014we}
\bibfield{author}{\bibinfo{person}{Magne J{\o}rgensen}.}
  \bibinfo{year}{2014}\natexlab{}.
\newblock \showarticletitle{What we do and don't know about software
  development effort estimation}.
\newblock \bibinfo{journal}{\emph{IEEE software}} \bibinfo{volume}{31},
  \bibinfo{number}{2} (\bibinfo{year}{2014}), \bibinfo{pages}{37--40}.
\newblock


\bibitem[Kabinna et~al\mbox{.}(2016)]%
        {kabinna2016logging}
\bibfield{author}{\bibinfo{person}{Suhas Kabinna}, \bibinfo{person}{Cor-Paul
  Bezemer}, \bibinfo{person}{Weiyi Shang}, {and} \bibinfo{person}{Ahmed~E
  Hassan}.} \bibinfo{year}{2016}\natexlab{}.
\newblock \showarticletitle{Logging library migrations: A case study for the
  apache software foundation projects}. In \bibinfo{booktitle}{\emph{2016
  IEEE/ACM 13th Working Conference on Mining Software Repositories (MSR)}}.
  IEEE, \bibinfo{pages}{154--164}.
\newblock


\bibitem[Katz(2020)]%
        {libraries.io}
\bibfield{author}{\bibinfo{person}{Jeremy Katz}.}
  \bibinfo{year}{2020}\natexlab{}.
\newblock \bibinfo{title}{Libraries.io Open Source Repository and Dependency
  Metadata}.
\newblock
\newblock
\urldef\tempurl%
\url{https://doi.org/10.5281/ZENODO.808272}
\showDOI{\tempurl}


\bibitem[Ketkar et~al\mbox{.}(2022)]%
        {ketkar2022inferring}
\bibfield{author}{\bibinfo{person}{Ameya Ketkar}, \bibinfo{person}{Oleg
  Smirnov}, \bibinfo{person}{Nikolaos Tsantalis}, \bibinfo{person}{Danny Dig},
  {and} \bibinfo{person}{Timofey Bryksin}.} \bibinfo{year}{2022}\natexlab{}.
\newblock \showarticletitle{Inferring and applying type changes}. In
  \bibinfo{booktitle}{\emph{Proceedings of the 44th International Conference on
  Software Engineering}}. \bibinfo{pages}{1206--1218}.
\newblock


\bibitem[Kitchenham and Pfleeger(2002)]%
        {KitchenhamSampling02}
\bibfield{author}{\bibinfo{person}{Barbara Kitchenham} {and}
  \bibinfo{person}{Shari~Lawrence Pfleeger}.} \bibinfo{year}{2002}\natexlab{}.
\newblock \showarticletitle{Principles of Survey Research: Part 5: Populations
  and Samples}.
\newblock \bibinfo{journal}{\emph{SIGSOFT Softw. Eng. Notes}}
  \bibinfo{volume}{27}, \bibinfo{number}{5} (\bibinfo{date}{sep}
  \bibinfo{year}{2002}), \bibinfo{pages}{17–20}.
\newblock
\showISSN{0163-5948}
\urldef\tempurl%
\url{https://doi.org/10.1145/571681.571686}
\showDOI{\tempurl}


\bibitem[Krippendorff(2013)]%
        {krippendorff2018content}
\bibfield{author}{\bibinfo{person}{Klaus Krippendorff}.}
  \bibinfo{year}{2013}\natexlab{}.
\newblock \bibinfo{booktitle}{\emph{Content analysis: An introduction to its
  methodology} (\bibinfo{edition}{3rd} ed.)}.
\newblock \bibinfo{publisher}{Sage publications}, \bibinfo{address}{Thousand
  Oaks, California}. 221--250 pages.
\newblock


\bibitem[Kula et~al\mbox{.}(2018)]%
        {kula2018developers}
\bibfield{author}{\bibinfo{person}{Raula~Gaikovina Kula},
  \bibinfo{person}{Daniel~M German}, \bibinfo{person}{Ali Ouni},
  \bibinfo{person}{Takashi Ishio}, {and} \bibinfo{person}{Katsuro Inoue}.}
  \bibinfo{year}{2018}\natexlab{}.
\newblock \showarticletitle{Do developers update their library dependencies?}
\newblock \bibinfo{journal}{\emph{Empirical Software Engineering}}
  \bibinfo{volume}{23}, \bibinfo{number}{1} (\bibinfo{year}{2018}),
  \bibinfo{pages}{384--417}.
\newblock


\bibitem[Landis and Koch(1977)]%
        {landis1977measurement}
\bibfield{author}{\bibinfo{person}{J~Richard Landis} {and}
  \bibinfo{person}{Gary~G Koch}.} \bibinfo{year}{1977}\natexlab{}.
\newblock \showarticletitle{The measurement of observer agreement for
  categorical data}.
\newblock \bibinfo{journal}{\emph{biometrics}} (\bibinfo{year}{1977}),
  \bibinfo{pages}{159--174}.
\newblock


\bibitem[Larios~Vargas et~al\mbox{.}(2020)]%
        {selectingLibraries}
\bibfield{author}{\bibinfo{person}{Enrique Larios~Vargas},
  \bibinfo{person}{Maur{\'\i}cio Aniche}, \bibinfo{person}{Christoph Treude},
  \bibinfo{person}{Magiel Bruntink}, {and} \bibinfo{person}{Georgios Gousios}.}
  \bibinfo{year}{2020}\natexlab{}.
\newblock \showarticletitle{Selecting third-party libraries: The
  practitioners’ perspective}. In \bibinfo{booktitle}{\emph{Proceedings of
  the 28th ACM joint meeting on european software engineering conference and
  symposium on the foundations of software engineering}}.
  \bibinfo{pages}{245--256}.
\newblock


\bibitem[Mahmood et~al\mbox{.}(2020)]%
        {mahmood2020systematic}
\bibfield{author}{\bibinfo{person}{Yasir Mahmood}, \bibinfo{person}{Nazri
  Kama}, {and} \bibinfo{person}{Azri Azmi}.} \bibinfo{year}{2020}\natexlab{}.
\newblock \showarticletitle{A systematic review of studies on use case points
  and expert-based estimation of software development effort}.
\newblock \bibinfo{journal}{\emph{Journal of Software: Evolution and Process}}
  \bibinfo{volume}{32}, \bibinfo{number}{7} (\bibinfo{year}{2020}),
  \bibinfo{pages}{e2245}.
\newblock


\bibitem[McCabe(1976)]%
        {mccabe1976complexity}
\bibfield{author}{\bibinfo{person}{Thomas~J McCabe}.}
  \bibinfo{year}{1976}\natexlab{}.
\newblock \showarticletitle{A complexity measure}.
\newblock \bibinfo{journal}{\emph{IEEE Transactions on software Engineering}}
  \bibinfo{number}{4} (\bibinfo{year}{1976}), \bibinfo{pages}{308--320}.
\newblock


\bibitem[Mojica et~al\mbox{.}(2013)]%
        {softwareReuse}
\bibfield{author}{\bibinfo{person}{Israel~J Mojica}, \bibinfo{person}{Bram
  Adams}, \bibinfo{person}{Meiyappan Nagappan}, \bibinfo{person}{Steffen
  Dienst}, \bibinfo{person}{Thorsten Berger}, {and} \bibinfo{person}{Ahmed~E
  Hassan}.} \bibinfo{year}{2013}\natexlab{}.
\newblock \showarticletitle{A large-scale empirical study on software reuse in
  mobile apps}.
\newblock \bibinfo{journal}{\emph{IEEE software}} \bibinfo{volume}{31},
  \bibinfo{number}{2} (\bibinfo{year}{2013}), \bibinfo{pages}{78--86}.
\newblock


\bibitem[Nguyen et~al\mbox{.}(2016)]%
        {NguyenAPIMapping16}
\bibfield{author}{\bibinfo{person}{Trong~Duc Nguyen}, \bibinfo{person}{Anh~Tuan
  Nguyen}, {and} \bibinfo{person}{Tien~N. Nguyen}.}
  \bibinfo{year}{2016}\natexlab{}.
\newblock \showarticletitle{Mapping {API} Elements for Code Migration with
  Vector Representations}. In \bibinfo{booktitle}{\emph{Proceedings of the 38th
  International Conference on Software Engineering Companion}} (Austin, Texas)
  \emph{(\bibinfo{series}{ICSE '16})}. \bibinfo{publisher}{Association for
  Computing Machinery}, \bibinfo{address}{New York, NY, USA},
  \bibinfo{pages}{756–758}.
\newblock
\showISBNx{9781450342056}
\urldef\tempurl%
\url{https://doi.org/10.1145/2889160.2892661}
\showDOI{\tempurl}


\bibitem[Ni et~al\mbox{.}(2021)]%
        {ni2021soar}
\bibfield{author}{\bibinfo{person}{Ansong Ni}, \bibinfo{person}{Daniel Ramos},
  \bibinfo{person}{Aidan~ZH Yang}, \bibinfo{person}{In{\^e}s Lynce},
  \bibinfo{person}{Vasco Manquinho}, \bibinfo{person}{Ruben Martins}, {and}
  \bibinfo{person}{Claire Le~Goues}.} \bibinfo{year}{2021}\natexlab{}.
\newblock \showarticletitle{Soar: a synthesis approach for data science api
  refactoring}. In \bibinfo{booktitle}{\emph{2021 IEEE/ACM 43rd International
  Conference on Software Engineering (ICSE)}}. IEEE, \bibinfo{pages}{112--124}.
\newblock


\bibitem[Nielsen et~al\mbox{.}(2021)]%
        {nielsen2021semantic}
\bibfield{author}{\bibinfo{person}{Benjamin~Barslev Nielsen},
  \bibinfo{person}{Martin~Toldam Torp}, {and} \bibinfo{person}{Anders
  M{\o}ller}.} \bibinfo{year}{2021}\natexlab{}.
\newblock \showarticletitle{Semantic patches for adaptation of javascript
  programs to evolving libraries}. In \bibinfo{booktitle}{\emph{2021 IEEE/ACM
  43rd International Conference on Software Engineering (ICSE)}}. IEEE,
  \bibinfo{pages}{74--85}.
\newblock


\bibitem[OpenAI(2023)]%
        {gpt4}
\bibfield{author}{\bibinfo{person}{R OpenAI}.} \bibinfo{year}{2023}\natexlab{}.
\newblock \showarticletitle{GPT-4 technical report}.
\newblock \bibinfo{journal}{\emph{arXiv}} (\bibinfo{year}{2023}),
  \bibinfo{pages}{2303--08774}.
\newblock


\bibitem[Qualtrics({[n.\,d.]})]%
        {qualtricsSampling}
\bibfield{author}{\bibinfo{person}{Qualtrics}.}
  \bibinfo{year}{[n.\,d.]}\natexlab{}.
\newblock \bibinfo{title}{How to use stratified random sampling in 2023}.
\newblock
  \bibinfo{howpublished}{https://www.qualtrics.com/experience-management/research/stratified-random-sampling/}.
\newblock


\bibitem[Sawant et~al\mbox{.}(2019)]%
        {apiDeprecation}
\bibfield{author}{\bibinfo{person}{Anand~Ashok Sawant}, \bibinfo{person}{Romain
  Robbes}, {and} \bibinfo{person}{Alberto Bacchelli}.}
  \bibinfo{year}{2019}\natexlab{}.
\newblock \showarticletitle{To react, or not to react: Patterns of reaction to
  API deprecation}.
\newblock \bibinfo{journal}{\emph{Empirical Software Engineering}}
  \bibinfo{volume}{24}, \bibinfo{number}{6} (\bibinfo{year}{2019}),
  \bibinfo{pages}{3824--3870}.
\newblock


\bibitem[Seaman(1999)]%
        {qualitativeMethods}
\bibfield{author}{\bibinfo{person}{Carolyn~B. Seaman}.}
  \bibinfo{year}{1999}\natexlab{}.
\newblock \showarticletitle{Qualitative methods in empirical studies of
  software engineering}.
\newblock \bibinfo{journal}{\emph{IEEE Transactions on software engineering}}
  \bibinfo{volume}{25}, \bibinfo{number}{4} (\bibinfo{year}{1999}),
  \bibinfo{pages}{557--572}.
\newblock


\bibitem[Sharma and Kushwaha(2012)]%
        {sharma2012estimation}
\bibfield{author}{\bibinfo{person}{Ashish Sharma} {and}
  \bibinfo{person}{Dharmender~Singh Kushwaha}.}
  \bibinfo{year}{2012}\natexlab{}.
\newblock \showarticletitle{Estimation of software development effort from
  requirements based complexity}.
\newblock \bibinfo{journal}{\emph{Procedia Technology}}  \bibinfo{volume}{4}
  (\bibinfo{year}{2012}), \bibinfo{pages}{716--722}.
\newblock


\bibitem[Teyton et~al\mbox{.}(2012)]%
        {teyton2012mining}
\bibfield{author}{\bibinfo{person}{Cedric Teyton}, \bibinfo{person}{Jean-Remy
  Falleri}, {and} \bibinfo{person}{Xavier Blanc}.}
  \bibinfo{year}{2012}\natexlab{}.
\newblock \showarticletitle{Mining library migration graphs}. In
  \bibinfo{booktitle}{\emph{2012 19th Working Conference on Reverse
  Engineering}}. IEEE, \bibinfo{pages}{289--298}.
\newblock


\bibitem[Teyton et~al\mbox{.}(2013)]%
        {teyton2013automatic}
\bibfield{author}{\bibinfo{person}{C{\'e}dric Teyton},
  \bibinfo{person}{Jean-R{\'e}my Falleri}, {and} \bibinfo{person}{Xavier
  Blanc}.} \bibinfo{year}{2013}\natexlab{}.
\newblock \showarticletitle{Automatic discovery of function mappings between
  similar libraries}. In \bibinfo{booktitle}{\emph{2013 20th Working Conference
  on Reverse Engineering (WCRE)}}. IEEE, \bibinfo{pages}{192--201}.
\newblock


\bibitem[Teyton et~al\mbox{.}(2014)]%
        {teyton2014study}
\bibfield{author}{\bibinfo{person}{C{\'e}dric Teyton},
  \bibinfo{person}{Jean-R{\'e}my Falleri}, \bibinfo{person}{Marc Palyart},
  {and} \bibinfo{person}{Xavier Blanc}.} \bibinfo{year}{2014}\natexlab{}.
\newblock \showarticletitle{A study of library migrations in java}.
\newblock \bibinfo{journal}{\emph{Journal of Software: Evolution and Process}}
  \bibinfo{volume}{26}, \bibinfo{number}{11} (\bibinfo{year}{2014}),
  \bibinfo{pages}{1030--1052}.
\newblock


\bibitem[Wang et~al\mbox{.}(2020b)]%
        {deprecatedPythonAPIs}
\bibfield{author}{\bibinfo{person}{Jiawei Wang}, \bibinfo{person}{Li Li},
  \bibinfo{person}{Kui Liu}, {and} \bibinfo{person}{Haipeng Cai}.}
  \bibinfo{year}{2020}\natexlab{b}.
\newblock \showarticletitle{Exploring how deprecated python library apis are
  (not) handled}. In \bibinfo{booktitle}{\emph{Proceedings of the 28th acm
  joint meeting on european software engineering conference and symposium on
  the foundations of software engineering}}. \bibinfo{pages}{233--244}.
\newblock


\bibitem[Wang et~al\mbox{.}(2020a)]%
        {empiricalAPIUsage}
\bibfield{author}{\bibinfo{person}{Ying Wang}, \bibinfo{person}{Bihuan Chen},
  \bibinfo{person}{Kaifeng Huang}, \bibinfo{person}{Bowen Shi},
  \bibinfo{person}{Congying Xu}, \bibinfo{person}{Xin Peng},
  \bibinfo{person}{Yijian Wu}, {and} \bibinfo{person}{Yang Liu}.}
  \bibinfo{year}{2020}\natexlab{a}.
\newblock \showarticletitle{An empirical study of usages, updates and risks of
  third-party libraries in java projects}. In \bibinfo{booktitle}{\emph{2020
  IEEE International Conference on Software Maintenance and Evolution
  (ICSME)}}. IEEE, \bibinfo{pages}{35--45}.
\newblock


\bibitem[Xing and Stroulia(2007)]%
        {xing2007api}
\bibfield{author}{\bibinfo{person}{Zhenchang Xing} {and} \bibinfo{person}{Eleni
  Stroulia}.} \bibinfo{year}{2007}\natexlab{}.
\newblock \showarticletitle{API-evolution support with Diff-CatchUp}.
\newblock \bibinfo{journal}{\emph{IEEE Transactions on Software Engineering}}
  \bibinfo{volume}{33}, \bibinfo{number}{12} (\bibinfo{year}{2007}),
  \bibinfo{pages}{818--836}.
\newblock


\bibitem[Xu et~al\mbox{.}(2020)]%
        {reinventingTheWheels}
\bibfield{author}{\bibinfo{person}{Bowen Xu}, \bibinfo{person}{Le An},
  \bibinfo{person}{Ferdian Thung}, \bibinfo{person}{Foutse Khomh}, {and}
  \bibinfo{person}{David Lo}.} \bibinfo{year}{2020}\natexlab{}.
\newblock \showarticletitle{Why reinventing the wheels? An empirical study on
  library reuse and re-implementation}.
\newblock \bibinfo{journal}{\emph{Empirical Software Engineering}}
  \bibinfo{volume}{25}, \bibinfo{number}{1} (\bibinfo{year}{2020}),
  \bibinfo{pages}{755--789}.
\newblock


\bibitem[Zhang et~al\mbox{.}(2020)]%
        {zhang2020deep}
\bibfield{author}{\bibinfo{person}{Zejun Zhang}, \bibinfo{person}{Minxue Pan},
  \bibinfo{person}{Tian Zhang}, \bibinfo{person}{Xinyu Zhou}, {and}
  \bibinfo{person}{Xuandong Li}.} \bibinfo{year}{2020}\natexlab{}.
\newblock \showarticletitle{Deep-diving into documentation to develop improved
  java-to-swift api mapping}. In \bibinfo{booktitle}{\emph{Proceedings of the
  28th International Conference on Program Comprehension}}.
  \bibinfo{pages}{106--116}.
\newblock


\end{thebibliography}

\end{document}